\documentclass[reprint,amsmath,amssymb,aps,superscriptaddress,floatfix]{revtex4-2}
\usepackage{amsmath,amssymb,amsthm}
\usepackage{graphicx}
\usepackage{color}
\usepackage{dcolumn}
\usepackage{bm}
\usepackage{multirow}
\usepackage{braket}
\usepackage{xcolor}
\usepackage[draft]{changes}
\usepackage[normalem]{ulem}
\usepackage{cancel}
\usepackage{booktabs}
\usepackage{tikz}
\usetikzlibrary{arrows.meta}

\usepackage{hyperref}
\hypersetup{
    colorlinks=true,
    linkcolor=blue,
    filecolor=magenta,      
    urlcolor=blue,
    citecolor=blue,
    pdfpagemode=FullScreen,
}

\date{\today}
\newcommand{\Tc}{T_\mathrm{c}}
\newcommand{\dd}{\mathrm{d}}
\newcommand{\qea}{q_{\mathrm{EA}}}
\newcommand{\qeainfty}{q_{\mathrm{EA}}^\infty}
\newcommand{\ueainfty}{u_{\mathrm{EA}}^\infty}
\newcommand{\registeredmark}{\textsuperscript{\tiny\textregistered}}

\begin{document}

\title{Testing Replica Symmetry Breaking in the three-dimensional Edwards-Anderson model by using vanishing fields}

\author{Miguel Aguilar-Janita}
    \email[Corresponding author:]{ miguel.aguilarjanita@uniroma1.it}
  \affiliation{Departamento de F\'{\i}sica, Universidad de Extremadura, 06006 Badajoz, Spain}
  \affiliation{Dipartimento di Fisica, Sapienza Universit\`a di Roma,  00185 Rome, Italy}

\author{Silvio~Franz}
    \affiliation{Dipartimento di Matematica e Fisica, Universit\`a del Salento, 73100 Lecce, Italy and Istituto Nazionale di Fisica Nucleare, Sezione di Lecce, 73100 Lecce, Italy}
    \affiliation{Universit\'e Paris-Saclay, CNRS, LPTMS, 91405, Orsay, France}

\author{Victor~Martin-Mayor}
    \affiliation{Departamento de F\'\i{}sica Te\'orica, Universidad Complutense, 28040 Madrid, Spain}

\author{Javier~Moreno-Gordo}
    \affiliation{Departamento de Didáctica de las Ciencias Experimentales y de las Matemáticas, Universidad de Extremadura, 10004 Cáceres, Spain}
    \affiliation{Instituto de Biocomputaci\'on y F\'{\i}sica de Sistemas Complejos (BIFI), 50018 Zaragoza, Spain}

\author{Giorgio~Parisi}
    \affiliation{International Research Center of Complexity Sciences, Hangzhou International Innovation Institute, Beihang University, Hangzhou 311115, China}
     \affiliation{Dipartimento di Fisica, Sapienza Universit\`a di Roma,  00185 Rome, Italy}
    \affiliation{CNR-Nanotec, Rome Unit, and INFN, Sezione di Roma, 00185 Rome, Italy}

\author{Federico~Ricci-Tersenghi}
    \affiliation{Dipartimento di Fisica, Sapienza Universit\`a di Roma,  00185 Rome, Italy}
    \affiliation{CNR-Nanotec, Rome Unit, and INFN, Sezione di Roma, 00185 Rome, Italy}

\author{J.J.~Ruiz-Lorenzo}
    \affiliation{Departamento de F\'{\i}sica, Universidad de Extremadura, 06006 Badajoz, Spain}
    \affiliation{Instituto de Computaci\'on Cient\'{\i}fica Avanzada (ICCAEx), Universidad de Extremadura, 06006 Badajoz, Spain}

\date{\today}

\begin{abstract}
Replica Symmetry Breaking (RSB) theory for spin glasses, which is
exact in infinite dimensions, provides detailed predictions for
the behavior of these systems. One of these predictions is that vanishing
magnetic fields induce chaotic behavior. Here, \emph{vanishing} has a precise meaning: we refer to an externally applied magnetic field with intensity scaling as $1/\sqrt{N}$, where $N$ is the
number of spins in the system. RSB predicts that this chaotic behavior depends only on the probability density function of
the overlap, $P(q)$, as computed at zero field.

We proceed to test the RSB prediction in the 3D
Edwards-Anderson model, by first
computing the $P(q)$ numerically in the absence of a magnetic field. We then use
this function as input to analytically compute the statistics of
the states in the regime of a vanishing magnetic field, using a
representation based on the Bolthausen-Sznitman coalescent process (which is exact for the
mean-field
Sherrington-Kirkpatrick model). This mixed numerical-analytical approach allows us to determine
the prediction of RSB for the overlap distribution function measured in the presence of a small
vanishing field, as well as the increase in chaoticity as the
field strength grows. 

The above RSB predictions are quantitatively compared with the outcome of 
numerical simulations of the three-dimensional  Edwards-Anderson   in the presence of a small vanishing field, finding
very good agreement. This agreement provides strong support for the idea that three-dimensional
spin glass at zero magnetic field and low temperature is well described by RSB theory.

\end{abstract}

\maketitle

\section{Introduction}
\label{sec:introduction}

Materials exhibiting very slow dynamics, mainly induced by frustration
and disorder, are ubiquitous in nature. Among them, structural glasses
and spin glasses are particularly noteworthy~\cite{mydosh:93,dahlberg:25}.

The latter have served as a testing ground for the development of
highly sophisticated analytical theories. In particular, Parisi’s
solution of spin glasses within the mean-field approximation
(equivalent to infinite spatial dimension) displays remarkable and
surprising properties, such as the existence of an infinite number of
pure states organized according to an ultrametric structure~\cite{mezard:87}.

This theory, known as the Replica Symmetry Breaking (RSB) theory, was
originally developed using a finite number of copies of the system
(replicas), which were analytically continued to zero at the end of
the calculation.

A more modern approach to these concepts can be obtained through the
notion of multiple equilibria, which furthermore allows spin glasses
and other complex physical systems to be placed within a broader
framework of complex systems in Science and the Social Sciences,
including economics~\cite{parisi:23}.

The results obtained in the mean-field setting using the RSB theory were rigorously 
established several decades later, placing Parisi's theory on a rigorous mathematical 
footing. In particular, Guerra's interpolation method~\cite{guerra:03}, Talagrand's proof 
of the Parisi formula~\cite{talagrand:06}, and Panchenko's derivation of ultrametricity~\cite{panchenko:13}
completed the mathematical justification of the mean-field spin-glass picture 
predicted by the RSB framework.

In addition, the continuous ultrametric tree that provides the organization of pure
states in infinite dimensions was also placed on rigorous mathematical
grounds by Ruelle through his probability cascades~\cite{ruelle:87}, a framework that
was later further developed by Bolthausen and Sznitman~\cite{bolthausen:98}.

Despite the broad applicability of RSB theory to many areas of
Science, the relevance of this framework for the description of their
initial target, namely three-dimensional spin glasses, remains an open
problem. In particular, alternative theoretical pictures have been proposed
to describe finite-dimensional spin glasses, including the droplets
model~\cite{bray:87,fisher:86,fisher:88}, the Trivial-Not-Trivial scenario~\cite{krzakala:00,palassini:00}, and the chaotic pairs theory~\cite{newman:92,newman:96b,newman:97,newman:98}.

Numerical simulations of finite-dimensional spin 
glasses~\footnote{In this paper, finite-dimensional spin glasses mean spin glasses at a dimensionality above the lower critical dimension and below the upper critical one. One can assume 5/2 \cite{maiorano:18, bernaschi:26b} as the lower critical dimension and 6 \cite{aguilar-janita:24b, aguilar-janita:25} as the upper one, although there is no general consensus~\cite{fisher:88, newman:25}.}
have been employed to decide which of the theoretical pictures provides the most accurate description.
Most simulations to date are in favor of the RSB theory, which is exact in the
mean-field limit. Indeed, many of its predictions appear to hold in finite dimensions as well~\cite{marinari:00,martin-mayor:22}. Examples
include the property known as overlap equivalence and stochastic
stability. However, there is still no direct numerical evidence for the
existence of an ultrametric organization in finite dimensions \cite{janus:11,martin-mayor:22}. Since we will provide (indirect) evidence for the ultrametric organization of the states in three dimensions, let us now spend a few words on this problem.

The difficulties encountered in demonstrating ultrametricity in simulations at finite dimensions are somewhat surprising because, in the mean-field setting, deriving ultrametricity is relatively simple.
One
only needs to invoke the Ghirlanda–Guerra identities~\cite{ghirlanda:98} together with
self-averaging of the free energy. Moreover, the Ghirlanda–Guerra identities can
themselves be deduced by assuming stochastic stability (which is exact
in the mean-field limit) of the Hamiltonian under stochastic
perturbations consisting in the addition of random Hamiltonians with arbitrary $p$-spin
interactions.

As already mentioned, there is strong numerical evidence supporting
stochastic stability as a property of finite-dimensional spin glasses
as well, both in equilibrium \cite{martin-mayor:22} and out-of-equilibrium studies \cite{franz:98,marinari:00}. 
Consequently, the presence of ultrametricity in finite
dimensions appears highly plausible, despite the considerable
technical difficulties involved in detecting it through numerical
simulations \cite{franz2000ultrametricity, contucci:07b}.

It should be noted that in Ref. \cite{marinari:00} it was rigorously demonstrated that,
assuming both the Ghirlanda–Guerra identities and ultrametricity, one
recovers exactly the same ultrametric structure as in infinite
dimensions, namely that one quarter of all overlap triangles are
equilateral, while the remaining three quarters are isosceles.

In this work, we shall investigate the structure of pure states in
three-dimensional spin glasses through a mixed analytical–numerical
approach. A similar analysis was previously carried out in four
dimensions, where the results were found to be fully compatible with
the predictions of RSB theory \cite{aguilar-janita:24}.

The central idea {is that the Gibbs state of a given spin glass sample
is characterized} by an ultrametric tree which can be generated using
the Bolthausen–Sznitman {coalescent construction}~\cite{bolthausen:98}.
 
This construction produces { trees} of pure states possessing
exactly the same statistical properties as those found in the
infinite-dimensional theory. To endow this construction with
predictive power, it is necessary to specify the overlap probability
distribution. Here we shall employ the overlap distribution obtained
numerically from simulations of the three-dimensional spin glass. For
this reason, we refer to our method as a mixed analytical–numerical
technique.

Another important ingredient of our approach is the introduction of
perturbations through vanishing external fields, for example, fields
whose strength decreases as $1/\sqrt{N}$ in the thermodynamic
limit~\cite{aguilar-janita:24}. In mean-field at least, such perturbations have the remarkable property of modifying
only the relative statistical weights of the different pure states,
while leaving the internal properties of each state unchanged.

The generation of spin-glass trees, 
permitting the reconstruction of the overlap probability distribution,
allows us to investigate the chaotic properties of the system by comparing, {for each random sample}
the equilibrium state in the absence of external fields with that
obtained in the presence of a vanishing field. This approach also
makes it possible to compare the predictions of our mixed
analytical–numerical framework—where the only numerical input is the
overlap distribution measured in zero magnetic field—with the results
of direct numerical simulations performed in the presence of such
vanishing perturbations.

As previously observed in four-dimensional spin glasses and spin glasses in Bethe lattices~\cite{aguilar-janita:24}, we find an
excellent agreement between the predictions of the mixed
analytical–numerical approach and the results obtained from direct
numerical simulations. 

The paper is structured as follows. In Sec.~\ref{sec:model} we introduce the Edwards-Anderson model in a vanishing magnetic field,  define the observables used throughout this work, and describe the method, based on the ultrametric tree construction, used to obtain RSB predictions from the zero-field overlap distribution. In Sec.~\ref{sec:results} we compare these predictions with direct Monte Carlo simulations performed in a vanishing magnetic field. Finally, in Sec.~\ref{sec:conclusions} we summarize our conclusions. 
Additional material is presented in five appendices: details on the simulations (Appendix~\ref{app:num}), some analytical background on the problem (Appendices~\ref{app:ultra1} and \ref{app:ultra2}), consistency checks of the zero-field overlap distribution (Appendix~\ref{app:checkPq}) and the study of the dependence on the vanishing fields and the lattice size (Appendix~\ref{app:hf}).

\section{Methods}
\label{sec:model}
\subsection{Model and simulations}
We consider the Edwards-Anderson model in a single cubic lattice with linear size $L$ and periodic boundary conditions. The $N=L^3$ Ising spins, $\sigma_i=\pm 1$,  occupy the nodes of the lattice $i$, and interact through the Hamiltonian:
\begin{equation}
    \mathcal{H} = -\sum_{\braket{ij}}J_{ij}\sigma_i\sigma_j - \dfrac{h}{L^{3/2}}\sum_i \sigma_i \, ,
\end{equation}
where $\braket{ij}$ denotes lattice nearest-neighbors.  The couplings $J_{i,j}$ are independent and identically distributed stochastic variables, $J_{ij}=\pm J$ with equal probability. We choose energy units such that $J=1$. Each realization of the couplings $\lbrace J_{ij} \rbrace$ is named a \textit{sample} and remains constant throughout the simulation. Different realizations of the same sample $\lbrace J_{ij} \rbrace_\alpha$ evolving with different thermal noise are named \textit{replicas}.
The externally applied magnetic field is $H=h L^{-3/2}$, where $h$ is the strength of the vanishing magnetic field~\cite{aguilar-janita:24}. At $h=0$, the system undergoes
a second-order phase transition at the critical temperature $\Tc = 1.1019(29)$~\cite{janus:13}.

Our Parallel tempering simulations have equilibrated 2560 samples with linear sizes $L=8,12,$ and 16, for temperatures $T\geq T_\text{min}=0.73\, \Tc$ (we present only the results obtained for $T_\text{min}$). We have used the thermalization test explained in Ref.~\cite{billoire:18}. The strengths of the vanishing fields have been $h=\lbrace 0,1,2,3,4,6,8,10\rbrace$. A more detailed description of our simulations can be found in Appendix \ref{app:num}.

\subsection{Observables}

We shall focus on $P(q)$, namely the probability density function for the
overlap between two different replicas, $\sigma$ and $\tau$
\begin{equation}
    P(q)=\overline{\,\Big\langle\delta\Big(q \ -\ \dfrac{1}{L^3}\sum_i\sigma_i\tau_i\Big)\Big\rangle}
\end{equation}
where $\delta(x)$ is Dirac's delta function, $\langle\ldots \rangle$ denotes the thermal average computed in a single sample, and the overline indicates the average over samples (always performed after the thermal average).

Specifically, we are interested in three different $P(q)$ functions~\footnote{Note that we do not introduce any binning in $q$ for the computation of $P(q)$; instead, we consider all the $2N+1$ possible values of $q$.}(the superscript $L$ will indicate the system size for which these $P(q)$ are computed). 

First, we simply denote $P^L(q)$ as the probability density function of the overlap computed with both replicas in the absence of an external magnetic field ($h=0$). 

Second, if one replica is at $h=0$ and the other is at $h \neq 0$, then we denote the probability density function of the overlap as $P^L_\text{C}(\lvert q \rvert ; h)$. The subscript $C$ stands for \emph{chaos} because $P^L_\text{C}(\lvert q \rvert ; h)$  quantifies the variations induced by the vanishing field. Notice that both
$P^L$ and $P^L_\text{C}$ are, by definition, even functions of $q$.

Finally, we write $P^L(q;h)$ if both replicas are subjected to the same $h>0$.

\begin{figure}
    \centering \includegraphics[width=0.9\linewidth]{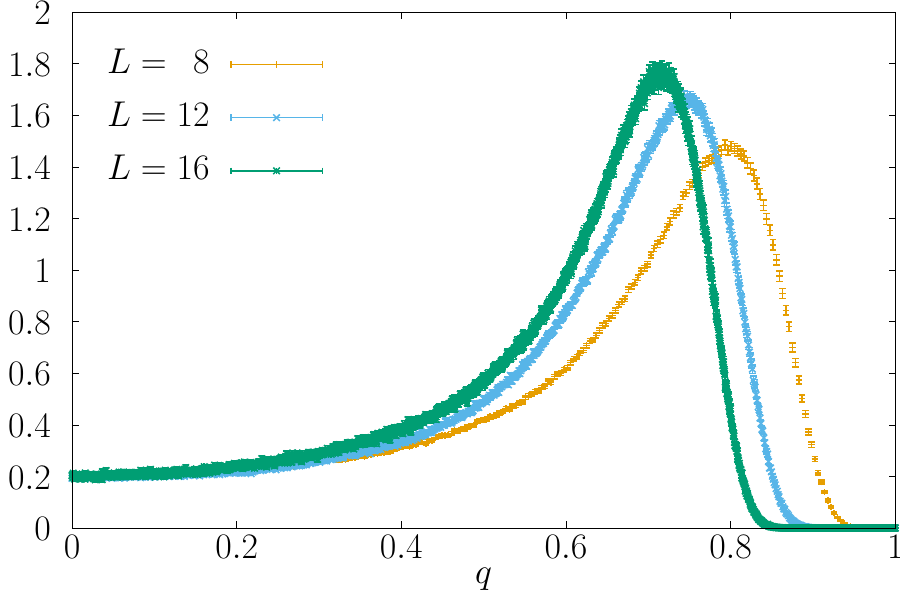}
    \caption{Shift of the probability $P(|q|)$ for different sizes of the system $L$ in three-dimensional EA model at $T=0.73 \Tc$ and $h=0$.}
    \label{fig:shift_peak}
\end{figure}

Let us emphasize that the size dependence of $P^L(q)$
is strong in both the height and position of the peaks; see Fig.~\ref{fig:shift_peak}, which makes extrapolating $P^L(q)$ to $L\to\infty$ difficult and noisy. Therefore, following~\cite{aguilar-janita:24}, we minimize those finite volume effects by studying the observables~\footnote{To generate a clear comparison, we have divided the numerical $P^L(q);0$ into a given number of equal-probability intervals.}:
\begin{equation}
    R^L_\text{C}(|q|;h) = \dfrac{P^L_\text{C}(|q|;h)}{P^L(|q|;0)}\, , \>\>\>\> R^L (q;h) = \dfrac{P^L(q;h)}{P^L(q;0)}\, , \label{eq:R_q_h}
\end{equation}
to complement the comparison of the probability density functions themselves~\footnote{To minimize systematic errors, we used the same samples to compute the observables in the numerator and the denominator.}.

\subsection{Ultrametric tree construction}
\label{sec:methods}
In this work, we use the ultrametric tree construction introduced in Ref.~\cite{aguilar-janita:24} to quantitatively predict the overlap distribution in the presence of a small external magnetic field in our three-dimensional spin glass. We follow a procedure similar to that in the four-dimensional case~\cite{aguilar-janita:24} (see Appendices \ref{app:ultra1} and \ref{app:ultra2} for implementation details).

First, we compute the probability density function of the overlaps in the absence of any external magnetic field, $P^L(q)$. This $P^L(q)$ and $q_{\mathrm{EA}}$ will be inputs to the ultrametric tree method.

To use $P^L(q)$ as input, we must model it with the structure of $P(q)$ in the thermodynamic limit according to RSB. Our ansatz models the central part of $P^L(q)$ with a quartic polynomial in $q$.
Because of the symmetry $P^L(q)=P^L(-q)$, only the even-order polynomial coefficients can be nonzero. We determine these coefficients by fitting the (symmetrized) numerical data for $q\ge 0$, $2 P^L(q)$, to an even polynomial over a small-$q$ range.

On the other hand,  our ansatz substitutes the  finite-size peak regions  of the $P^L(q)$  in Fig. ~\ref{fig:shift_peak} with Dirac's delta functions centered at $\pm q_{\mathrm{EA}}$, and the  symmetrized numerical data can be fitted as
\begin{eqnarray}
    P_{\mathrm{fit}}(\lvert q \rvert)&=&\Theta(q_{\mathrm{EA}} - \lvert q \rvert)\left[ a+bq^2+cq^4 \right] \label{eq:Pq_model}\\ 
    &+& \big(1-x_\mathrm{M}\big) \delta(\lvert q\rvert - q_{\mathrm{EA}})\, , \nonumber\\[0.2mm]
    x_\mathrm{M}&=&\int_0^{q_\mathrm{EA}}\,\mathrm{d}q\, (a+bq^2+cq^4)\,,
\end{eqnarray}
where $a$, $b$, and $c$ are the non-zero coefficients of  the quartic polynomial, $\Theta$ is Heaviside' step function and $\delta$ is Dirac's delta function.

The polynomial coefficients, obtained from a fit limited to $0\le q <0.5$ for $L=8$, $0 \le q  <0.45$ for $L=12$, and $0 \le q <0.4$ for $L=16$, are in Table~\ref{tab:ajuste_pq_cuadratica}. As for
the center of the Dirac's delta functions, $\qea$, it is determined by imposing that the model
and the numerical data have the same average value for $\vert q\vert$:
\begin{equation}\label{eq:matching-model-dirac}
\int_0^{\qea}\, \mathrm{d}q \, (aq+bq^3+cq^5)\ +\ [1-x_\mathrm{M}]\qea =\overline{\langle\vert q\vert\rangle} 
\end{equation}
Note that the l.h.s. of Eq.~\eqref{eq:matching-model-dirac} vanishes for $\qea=0$ and grows monotonically with $\qea$. 
Thus, Eq.~\eqref{eq:matching-model-dirac} has a unique solution for $\qea<1$.

\begin{table}[htpb]
    \centering
    \setlength{\tabcolsep}{6pt}
    \begin{tabular}{l c c c c c c}
        \hline \hline
        $L$ & $a$ & $b$ & $c$ & $q_{\mathrm{EA}}$ & $\chi^2/\text{d.o.f.}$ & $p$-value\\
        \hline
        8  & 0.414 & 0.781 & 3.464 & 0.807 & 54.40/126 & 1.00 \\
        12 & 0.395 & 0.865 & 5.365 & 0.750 & 128.44/386 & 1.00 \\
        16 & 0.404 & 1.593 & 4.080 & 0.714 & 440.58/817 & 1.00 \\
        \hline \hline
    \end{tabular}
    \caption{Fitted polynomial coefficients, see Eq.~(\ref{eq:Pq_model}) for definition, as obtained for all our simulated system sizes. Data at different $q$ are extremely correlated; in particular, the data chosen near the minimum to avoid the effects of the peaks explain the  small values of $\chi^2$ as compared to the number of degrees of freedom. and a $p$-value near 1.}
    \label{tab:ajuste_pq_cuadratica}
\end{table}

\begin{figure}
    \centering
    \includegraphics[width=0.9\linewidth]{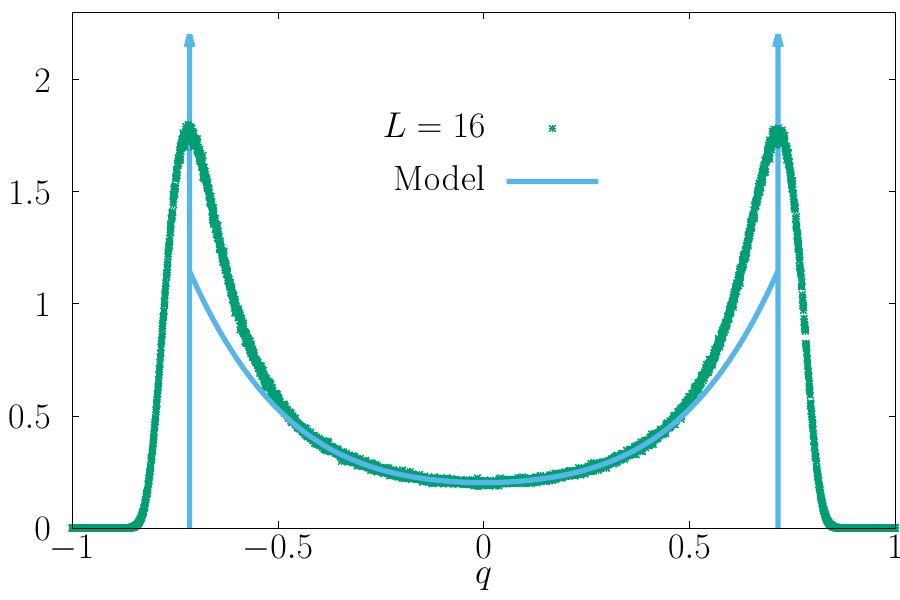}
    \caption{Scheme of the modeling of $P(q)$ computed in numerical simulations at zero magnetic field at $T=0.73 \Tc$.}
    \label{fig:scheme_Pq}
\end{figure}

For each value of $L$ we have used for building their $P_\text{fit}$ the values reported in Table \ref{tab:ajuste_pq_cuadratica}. Fig.~\ref{fig:scheme_Pq} illustrates our modeling scheme for $P^L(q)$. Further tests of this representation of the numerical $P^L(q)$ with its proxy $P_\text{fit}(q)$ are presented in Appendix \ref{app:checkPq}. 

Determining the $P_\mathrm{fit}(q)$ is only a crucial first step for the application of the ultrametric-tree method,  however. What one really needs is the inverse of the cumulative distribution function for $q\ge 0$
\begin{equation}\label{eq:x-q-def}
    x( q )=\int_0^{q}\mathrm{d} q  \, P_\mathrm{fit}(q )\,
\end{equation}
We numerically compute the inverse $q(x)$ simply by tabulating $x(q)$ over a dense grid of values and reading the table in reverse order. We extend the
tabulated $q(x)$ to a continuous function by fitting it to the function~\footnote{This functional form is a Padé approximant for the odd function $q(x)$. Note that $q(x)$ is an odd function because also $x(q)$ is an odd function, see Eq.~\eqref{eq:x-q-def}.} 
\begin{equation}
    q(x)=\dfrac{x(a_1+b_1x^2)}{1+c_1x^2} \, . \label{eq:qx}
\end{equation}
The values of $a_1$, $b_1$, and $c_1$ are the inputs used in the ultrametric tree method. We need the continuous version of $q(x)$ to obtain the overlap between different leaves of the tree. The overlap is the value of the function $q(x)$ evaluated at the merging point of the branches to which the leaves belong.

The last ingredient needed by the ultrametric tree method is the choice of the size of the trees (i.e., the number of 
leaves in a tree) and of the number of different realizations of the trees. Indeed, because the weights associated to the leaves 
are themselves random variables, a single function $q(x)$ may generate an infinite number of different trees.
Several numerical experiments have taught us that the output of the computation becomes independent of these two 
choices if we set  $10^5$ leaves (or larger) for the size of the tree  and  $\sim 10^5$ (or larger) for the number of 
trees generated.

\section{Results and Discussion }
\label{sec:results}

\begin{figure}
    \centering
    \includegraphics[width=0.99\linewidth]{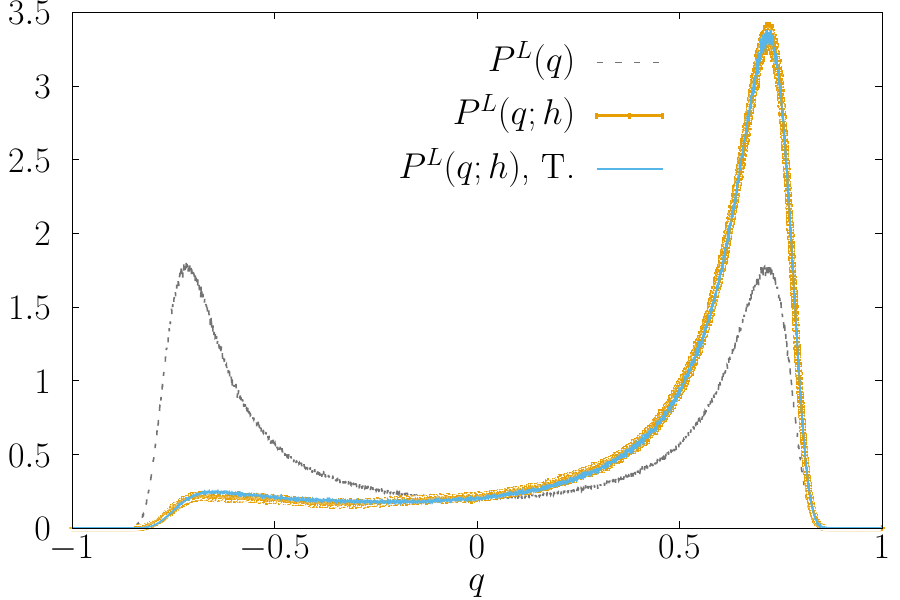}
    \caption{The functions $P^L(q)$ and $P^L(q;h)$ as a function of $q$ at $T=0.73T_c$ for $h=3$ for the largest lattice size simulated, $L=16$. The agreement between the numerical $P^L(q;h)$ and its theoretical prediction obtained from the ultrametric tree (denoted by T.) is very good. The largest deviation occurs in the negative peak, which is a finite-size effect. The theoretical curve has been obtained as the product of $R^L(q;h)$ and the numerical zero-field probability distribution of overlaps, $P^L(q)$ [see Eq.~\eqref{eq:R_q_h}].}
    \label{fig:Pq_straight}
\end{figure}

\begin{figure}
    \centering
    \includegraphics[width=0.99\linewidth]{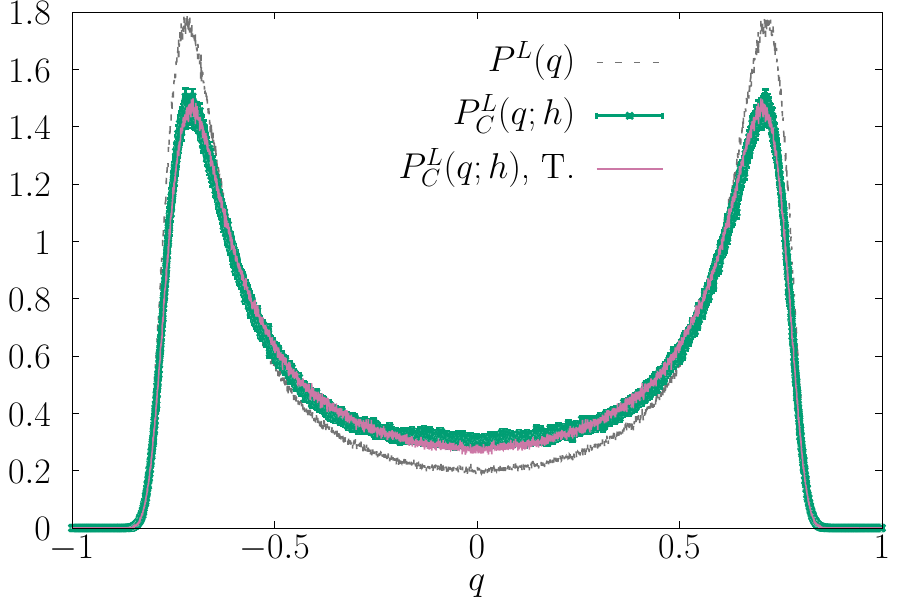}
    \caption{The functions $P^L(q)$ and $P^L_\text{C}(q;h)$ as a function of $q$ at $T=0.73T_c$ and for $h=3$ for $L=16$ (our largest lattice). The numerical curve agrees well with the theoretical prediction from the ultrametric tree (denoted by T.). This curve have been obtained as the product of $R^L_\text{C}(q;h)$ and the numerical zero-field probability distribution of overlaps, $P^L(q)$ [see Eq.~\eqref{eq:R_q_h}].}
    \label{fig:Pq_chaos}
\end{figure}

Directly comparing the $P(q)$ obtained from the ultrametric tree with the numerical data is somewhat tricky because we used delta functions to generate the trees; recall Fig.~\ref{fig:scheme_Pq}. We have bypassed this problem following a two-step procedure. First, we determined the probability ratios in Eq.~\eqref{eq:R_q_h} from the ultrametric trees (Appendix~\ref{app:ultra2}). Second, we multiplied the probability ratios by the numerical $P^L(q)$ to obtain reasonably well-rounded peaks (the delta functions in the infinite-volume limit are broadened by finite-size effects). When employed for $h=0.3$ and our largest lattice $L=16$, we obtain a
nice agreement between the numerical results and the ultrametric-tree computations for $P^L(q;h)$ (see Fig.~\ref{fig:Pq_straight}) and for $P^L_C(q;h)$ (see Fig.~\ref{fig:Pq_chaos}). The theoretical prediction correctly captures the significant change in peak height in $P^L(q;h)$ (see Fig.~\ref{fig:Pq_straight}) and the slight redistribution of probability from the peaks to the region near $q=0$ in $P^L_C(q;h)$ (see Fig.~\ref{fig:Pq_chaos}).

\begin{figure}
    \centering
    \includegraphics[width=0.99\linewidth]{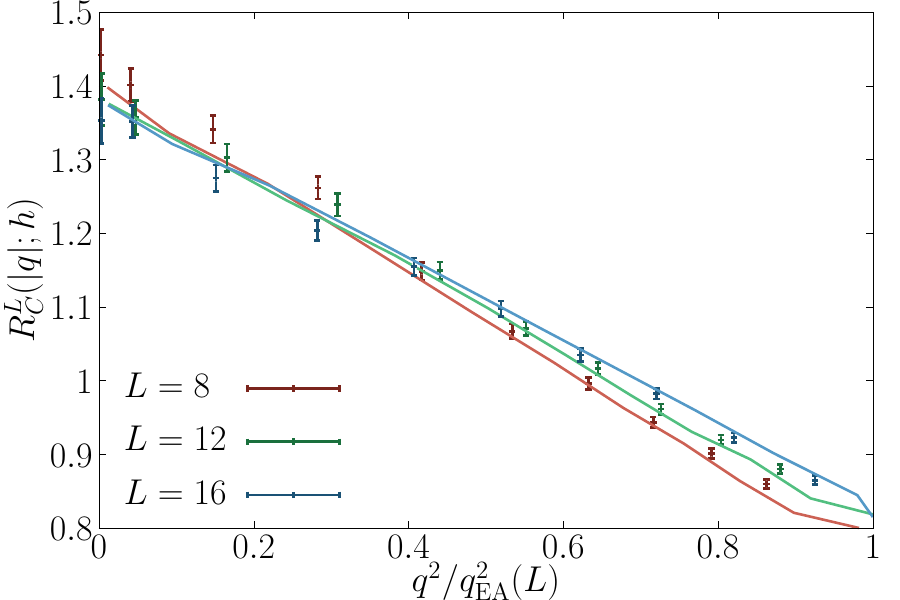}
    \caption{The ratio $R^L_C(q;h)$ as a function of $q^2/\qea^2(L)$ at $T=0.73T_c$ for $h=3$. Lines correspond to the theoretical prediction from the ultrametric tree, while points with errors correspond to the results from numerical simulations. }
    \label{fig:RC}
\end{figure}

\begin{figure}
    \centering
    \includegraphics[width=0.99\linewidth]{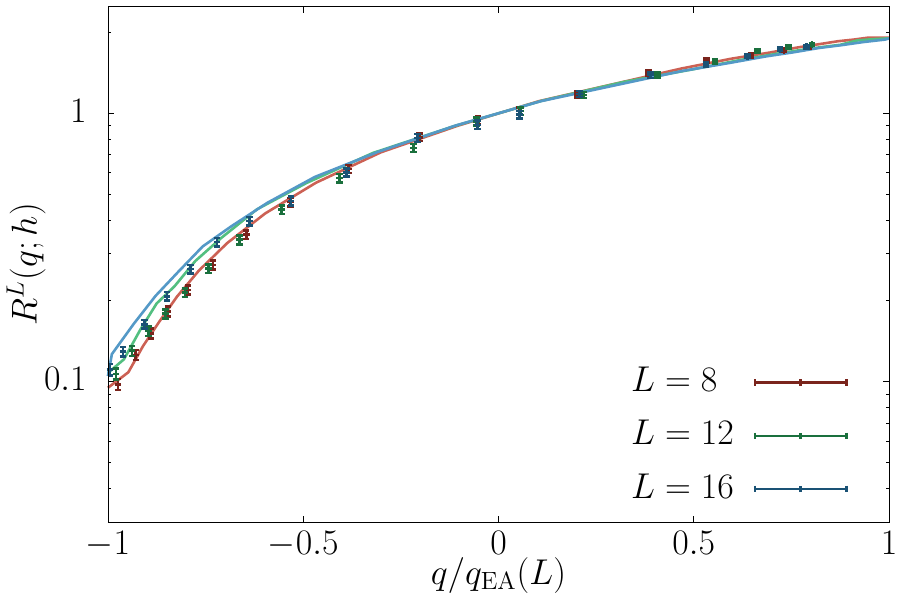}
    \caption{The ratio $R^L(q;h)$ as a function of $q/\qea(L)$ at $T=0.73T_c$ for $h=3$. Lines show the theoretical prediction from the ultrametric tree, while points with errors show results from numerical simulations.}
    \label{fig:RS}
\end{figure}

To properly appreciate the quality of the comparison between numerical data and theory, we plot the ratios $R^L_C (|q|;h)$ and $R^L (|q|;h)$ defined in Eq.~\eqref{eq:R_q_h} (see Fig.~\ref{fig:RC} and Fig.~\ref{fig:RS}, respectively). We only plot $q \in [ -\qea,\qea ]$, which is the range where the comparison between the analytic prediction and the data is feasible.

We find that the quality of the predictions starts to deteriorate for greater values of $h$. The particular value of $h$ at which the theoretical prediction stops being compatible with the numerical results depends both on $L$ and on whether one is looking at the chaotic or non-chaotic case. In Appendix \ref{app:hf} we present a detailed study comparing different lattice sizes for different values of $h$, up to $h=8$.

Analogously to the four-dimensional case discussed in Ref.~\cite{aguilar-janita:24}, the ratio $R^L_C (|q|;h)$ can, to a good approximation, be modeled as a linear function of $q^2$. On the other hand, a clear
difference appears with the four-dimensional case: as $q^2$ increases the errors diminish  and the size-dependence for both ratios $R^L_C (|q|;h)$ and $R^L (|q|;h)$ decrease.

\section{Conclusions}
\label{sec:conclusions}
We studied the behavior of the three-dimensional Edwards-Anderson model in a vanishing magnetic field, namely one whose strength decreases with the number of spins as $1/\sqrt{N}$.
In this regime, one of the predictions of RSB theory is that even such a small perturbation changes the relative weights of the equilibrium states, producing an $\mathcal{O}(1)$ effect. Given that the effect is sizeable, numerical simulations allow us to test some of the predictions of RSB theory in a three-dimensional setting. For larger magnetic fields, such a detailed
extrapolation from zero-field is not yet feasible as it would require further analytical progress  (because the properties of the equilibrium states themselves are expected to change with respect to their zero-field properties).

We have used the zero-field overlap probability distribution $P(q)$ as the only input for the computation. We then process it analytically through the Bolthausen-Sznitman coalescent process in order to obtain predictions for small magnetic fields, in particular for the two overlap probability distributions and related observables. We find very good agreement between this mixed numerical-analytical approach and direct numerical simulations performed in the presence of the corresponding magnetic fields. Therefore, the results presented in this work support describing the spin-glass phase of the three-dimensional Ising spin glass in terms of RSB theory.

We would like to emphasize that this analysis provides support for 
the presence of an ultrametric organization of the equilibrium 
states of the 3D Edwards-Anderson model, with the same structure as 
in mean-field theory. This is because the ultrametric properties are 
encoded in the Bolthausen-Sznitman coalescent process. As emphasized 
in the Introduction, obtaining traces of ultrametricity has been 
extremely elusive in three-dimensional studies, up to now. Our 
computation is a clear step forward in this direction.

\begin{acknowledgments}
This work was partially supported by the Ministerio de Ciencia, Innovación y Universidades (Spain) and by the European Regional Development Fund through grants no. 
 PID2022-136374NB-C21, PID2024-156352NB-I00, PID2024-158623NB-C22 and RED2022-134244-T (MCIU/AEI/10.13039/501100011033/FEDER, UE); by Junta de Ex\-tre\-ma\-dura (Spain) through grant no.~GR24022. We acknowledge the use of the computer facilities of the Instituto de Computacion Cient\'{\i}fica Avanzada de Extremadura (ICCAEx).
\end{acknowledgments}
\section*{Data and Code Availability}
The data supporting the ﬁndings of this study, and the simulation and analysis codes, are available from the corresponding author upon reasonable request.
\section*{Author Contributions}
M.A.-J. and J.M.-G., performed the numerical simulations;
M.A.-J., S.F., V.M.-M., J.M.-G., G.P., F.R.-T., and J.J.R.-L. analyzed data; 
and M.A.-J., S.F., V.M.-M., J.M.-G., G.P., F.R.-T., and J.J.R.-L. wrote the paper.
\section*{Competing Interests}
The authors declare no competing interests.
\bibliography{biblio}

\begin{thebibliography}{44}%
\makeatletter
\providecommand \@ifxundefined [1]{%
 \@ifx{#1\undefined}
}%
\providecommand \@ifnum [1]{%
 \ifnum #1\expandafter \@firstoftwo
 \else \expandafter \@secondoftwo
 \fi
}%
\providecommand \@ifx [1]{%
 \ifx #1\expandafter \@firstoftwo
 \else \expandafter \@secondoftwo
 \fi
}%
\providecommand \natexlab [1]{#1}%
\providecommand \enquote  [1]{``#1''}%
\providecommand \bibnamefont  [1]{#1}%
\providecommand \bibfnamefont [1]{#1}%
\providecommand \citenamefont [1]{#1}%
\providecommand \href@noop [0]{\@secondoftwo}%
\providecommand \href [0]{\begingroup \@sanitize@url \@href}%
\providecommand \@href[1]{\@@startlink{#1}\@@href}%
\providecommand \@@href[1]{\endgroup#1\@@endlink}%
\providecommand \@sanitize@url [0]{\catcode `\\12\catcode `\$12\catcode `\&12\catcode `\#12\catcode `\^12\catcode `\_12\catcode `\%12\relax}%
\providecommand \@@startlink[1]{}%
\providecommand \@@endlink[0]{}%
\providecommand \url  [0]{\begingroup\@sanitize@url \@url }%
\providecommand \@url [1]{\endgroup\@href {#1}{\urlprefix }}%
\providecommand \urlprefix  [0]{URL }%
\providecommand \Eprint [0]{\href }%
\providecommand \doibase [0]{https://doi.org/}%
\providecommand \selectlanguage [0]{\@gobble}%
\providecommand \bibinfo  [0]{\@secondoftwo}%
\providecommand \bibfield  [0]{\@secondoftwo}%
\providecommand \translation [1]{[#1]}%
\providecommand \BibitemOpen [0]{}%
\providecommand \bibitemStop [0]{}%
\providecommand \bibitemNoStop [0]{.\EOS\space}%
\providecommand \EOS [0]{\spacefactor3000\relax}%
\providecommand \BibitemShut  [1]{\csname bibitem#1\endcsname}%
\let\auto@bib@innerbib\@empty
\bibitem [{\citenamefont {Mydosh}(1993)}]{mydosh:93}%
  \BibitemOpen
  \bibfield  {author} {\bibinfo {author} {\bibfnamefont {J.~A.}\ \bibnamefont {Mydosh}},\ }\href@noop {} {\emph {\bibinfo {title} {Spin Glasses: an Experimental Introduction}}}\ (\bibinfo  {publisher} {Taylor and Francis},\ \bibinfo {address} {London},\ \bibinfo {year} {1993})\BibitemShut {NoStop}%
\bibitem [{\citenamefont {Dahlberg}\ \emph {et~al.}(2025)\citenamefont {Dahlberg}, \citenamefont {González-Adalid~Pemartín}, \citenamefont {Marinari}, \citenamefont {Martin-Mayor}, \citenamefont {Moreno-Gordo}, \citenamefont {Orbach}, \citenamefont {Paga}, \citenamefont {Parisi}, \citenamefont {Ricci-Tersenghi}, \citenamefont {Ruiz-Lorenzo},\ and\ \citenamefont {Yllanes}}]{dahlberg:25}%
  \BibitemOpen
  \bibfield  {author} {\bibinfo {author} {\bibfnamefont {E.}~\bibnamefont {Dahlberg}}, \bibinfo {author} {\bibfnamefont {I.}~\bibnamefont {González-Adalid~Pemartín}}, \bibinfo {author} {\bibfnamefont {E.}~\bibnamefont {Marinari}}, \bibinfo {author} {\bibfnamefont {V.}~\bibnamefont {Martin-Mayor}}, \bibinfo {author} {\bibfnamefont {J.}~\bibnamefont {Moreno-Gordo}}, \bibinfo {author} {\bibfnamefont {R.}~\bibnamefont {Orbach}}, \bibinfo {author} {\bibfnamefont {I.}~\bibnamefont {Paga}}, \bibinfo {author} {\bibfnamefont {G.}~\bibnamefont {Parisi}}, \bibinfo {author} {\bibfnamefont {F.}~\bibnamefont {Ricci-Tersenghi}}, \bibinfo {author} {\bibfnamefont {J.}~\bibnamefont {Ruiz-Lorenzo}},\ and\ \bibinfo {author} {\bibfnamefont {D.}~\bibnamefont {Yllanes}},\ }\bibfield  {title} {\bibinfo {title} {Spin-glass dynamics: experiment, theory and simulation},\ }\href {https://doi.org/10.1103/ctp2-zwyr} {\bibfield  {journal} {\bibinfo  {journal} {Rev. Mod. Phys.}\ }\textbf {\bibinfo {volume} {97}},\ \bibinfo {pages}
  {045005} (\bibinfo {year} {2025})}\BibitemShut {NoStop}%
\bibitem [{\citenamefont {M{\'e}zard}\ \emph {et~al.}(1987)\citenamefont {M{\'e}zard}, \citenamefont {Parisi},\ and\ \citenamefont {Virasoro}}]{mezard:87}%
  \BibitemOpen
  \bibfield  {author} {\bibinfo {author} {\bibfnamefont {M.}~\bibnamefont {M{\'e}zard}}, \bibinfo {author} {\bibfnamefont {G.}~\bibnamefont {Parisi}},\ and\ \bibinfo {author} {\bibfnamefont {M.}~\bibnamefont {Virasoro}},\ }\href {https://doi.org/10.1142/0271} {\emph {\bibinfo {title} {Spin-Glass Theory and Beyond}}}\ (\bibinfo  {publisher} {World Scientific},\ \bibinfo {address} {Singapore},\ \bibinfo {year} {1987})\BibitemShut {NoStop}%
\bibitem [{\citenamefont {Parisi}(2023)}]{parisi:23}%
  \BibitemOpen
  \bibfield  {author} {\bibinfo {author} {\bibfnamefont {G.}~\bibnamefont {Parisi}},\ }\bibfield  {title} {\bibinfo {title} {Nobel lecture: Multiple equilibria},\ }\href {https://doi.org/10.1103/RevModPhys.95.030501} {\bibfield  {journal} {\bibinfo  {journal} {Rev. Mod. Phys.}\ }\textbf {\bibinfo {volume} {95}},\ \bibinfo {pages} {030501} (\bibinfo {year} {2023})}\BibitemShut {NoStop}%
\bibitem [{\citenamefont {Guerra}(2003)}]{guerra:03}%
  \BibitemOpen
  \bibfield  {author} {\bibinfo {author} {\bibfnamefont {F.}~\bibnamefont {Guerra}},\ }\bibfield  {title} {\bibinfo {title} {Broken replica symmetry bounds in the mean field spin glass model},\ }\href {https://doi.org/10.1007/s00220-002-0773-5} {\bibfield  {journal} {\bibinfo  {journal} {Comm. Math. Phys.}\ }\textbf {\bibinfo {volume} {233}},\ \bibinfo {pages} {1} (\bibinfo {year} {2003})}\BibitemShut {NoStop}%
\bibitem [{\citenamefont {Talagrand}(2006)}]{talagrand:06}%
  \BibitemOpen
  \bibfield  {author} {\bibinfo {author} {\bibfnamefont {M.}~\bibnamefont {Talagrand}},\ }\bibfield  {title} {\bibinfo {title} {The {P}arisi formula},\ }\href {https://doi.org/10.4007/annals.2006.163.221} {\bibfield  {journal} {\bibinfo  {journal} {Ann. of Math.}\ }\textbf {\bibinfo {volume} {163}},\ \bibinfo {pages} {221} (\bibinfo {year} {2006})}\BibitemShut {NoStop}%
\bibitem [{\citenamefont {Panchenko}(2013)}]{panchenko:13}%
  \BibitemOpen
  \bibfield  {author} {\bibinfo {author} {\bibfnamefont {D.}~\bibnamefont {Panchenko}},\ }\bibfield  {title} {\bibinfo {title} {The {P}arisi ultrametricity conjecture},\ }\href {https://doi.org/10.4007/annals.2013.177.1.8} {\bibfield  {journal} {\bibinfo  {journal} {Ann. of Math.}\ }\textbf {\bibinfo {volume} {177}},\ \bibinfo {pages} {383} (\bibinfo {year} {2013})}\BibitemShut {NoStop}%
\bibitem [{\citenamefont {Ruelle}(1987)}]{ruelle:87}%
  \BibitemOpen
  \bibfield  {author} {\bibinfo {author} {\bibfnamefont {D.}~\bibnamefont {Ruelle}},\ }\bibfield  {title} {\bibinfo {title} {A mathematical reformulation of {D}errida's {REM} and {GREM}},\ }\href {https://doi.org/10.1007/BF01210613} {\bibfield  {journal} {\bibinfo  {journal} {Comm. Math. Phys.}\ }\textbf {\bibinfo {volume} {108}},\ \bibinfo {pages} {225} (\bibinfo {year} {1987})}\BibitemShut {NoStop}%
\bibitem [{\citenamefont {Bolthausen}\ and\ \citenamefont {Sznitman}(1998)}]{bolthausen:98}%
  \BibitemOpen
  \bibfield  {author} {\bibinfo {author} {\bibfnamefont {E.}~\bibnamefont {Bolthausen}}\ and\ \bibinfo {author} {\bibfnamefont {A.-S.}\ \bibnamefont {Sznitman}},\ }\bibfield  {title} {\bibinfo {title} {On {R}uelle's probability cascades and an abstract cavity method},\ }\href {https://doi.org/10.1007/s002200050450} {\bibfield  {journal} {\bibinfo  {journal} {Comm. Math. Phys.}\ }\textbf {\bibinfo {volume} {197}},\ \bibinfo {pages} {247} (\bibinfo {year} {1998})}\BibitemShut {NoStop}%
\bibitem [{\citenamefont {Bray}\ and\ \citenamefont {Moore}(1987)}]{bray:87}%
  \BibitemOpen
  \bibfield  {author} {\bibinfo {author} {\bibfnamefont {A.~J.}\ \bibnamefont {Bray}}\ and\ \bibinfo {author} {\bibfnamefont {M.~A.}\ \bibnamefont {Moore}},\ }\bibfield  {title} {\bibinfo {title} {Scaling theory of the ordered phase of spin glasses},\ }in\ \href@noop {} {\emph {\bibinfo {booktitle} {Heidelberg Colloquium on Glassy Dynamics}}},\ \bibinfo {series and number} {\bibinfo {series} {Lecture Notes in Physics}\ No.\ \bibinfo {number} {275}},\ \bibinfo {editor} {edited by\ \bibinfo {editor} {\bibfnamefont {J.~L.}\ \bibnamefont {van Hemmen}}\ and\ \bibinfo {editor} {\bibfnamefont {I.}~\bibnamefont {Morgenstern}}}\ (\bibinfo  {publisher} {Springer},\ \bibinfo {address} {Berlin},\ \bibinfo {year} {1987})\BibitemShut {NoStop}%
\bibitem [{\citenamefont {Fisher}\ and\ \citenamefont {Huse}(1986)}]{fisher:86}%
  \BibitemOpen
  \bibfield  {author} {\bibinfo {author} {\bibfnamefont {D.~S.}\ \bibnamefont {Fisher}}\ and\ \bibinfo {author} {\bibfnamefont {D.~A.}\ \bibnamefont {Huse}},\ }\bibfield  {title} {\bibinfo {title} {Ordered phase of short-range ising spin-glasses},\ }\href {https://doi.org/10.1103/PhysRevLett.56.1601} {\bibfield  {journal} {\bibinfo  {journal} {Phys. Rev. Lett.}\ }\textbf {\bibinfo {volume} {56}},\ \bibinfo {pages} {1601} (\bibinfo {year} {1986})}\BibitemShut {NoStop}%
\bibitem [{\citenamefont {Fisher}\ and\ \citenamefont {Huse}(1988)}]{fisher:88}%
  \BibitemOpen
  \bibfield  {author} {\bibinfo {author} {\bibfnamefont {D.~S.}\ \bibnamefont {Fisher}}\ and\ \bibinfo {author} {\bibfnamefont {D.~A.}\ \bibnamefont {Huse}},\ }\bibfield  {title} {\bibinfo {title} {Nonequilibrium dynamics of spin glasses},\ }\href {https://doi.org/10.1103/PhysRevB.38.373} {\bibfield  {journal} {\bibinfo  {journal} {Phys. Rev. B}\ }\textbf {\bibinfo {volume} {38}},\ \bibinfo {pages} {373} (\bibinfo {year} {1988})}\BibitemShut {NoStop}%
\bibitem [{\citenamefont {Krzakala}\ and\ \citenamefont {Martin}(2000)}]{krzakala:00}%
  \BibitemOpen
  \bibfield  {author} {\bibinfo {author} {\bibfnamefont {F.}~\bibnamefont {Krzakala}}\ and\ \bibinfo {author} {\bibfnamefont {O.~C.}\ \bibnamefont {Martin}},\ }\bibfield  {title} {\bibinfo {title} {Spin and link overlaps in three-dimensional spin glasses},\ }\href {https://doi.org/10.1103/PhysRevLett.85.3013} {\bibfield  {journal} {\bibinfo  {journal} {Phys. Rev. Lett.}\ }\textbf {\bibinfo {volume} {85}},\ \bibinfo {pages} {3013} (\bibinfo {year} {2000})}\BibitemShut {NoStop}%
\bibitem [{\citenamefont {Palassini}\ and\ \citenamefont {Young}(2000)}]{palassini:00}%
  \BibitemOpen
  \bibfield  {author} {\bibinfo {author} {\bibfnamefont {M.}~\bibnamefont {Palassini}}\ and\ \bibinfo {author} {\bibfnamefont {A.~P.}\ \bibnamefont {Young}},\ }\bibfield  {title} {\bibinfo {title} {Nature of the spin glass state},\ }\href {https://doi.org/10.1103/PhysRevLett.85.3017} {\bibfield  {journal} {\bibinfo  {journal} {Phys. Rev. Lett.}\ }\textbf {\bibinfo {volume} {85}},\ \bibinfo {pages} {3017} (\bibinfo {year} {2000})}\BibitemShut {NoStop}%
\bibitem [{\citenamefont {Newman}\ and\ \citenamefont {Stein}(1992)}]{newman:92}%
  \BibitemOpen
  \bibfield  {author} {\bibinfo {author} {\bibfnamefont {C.~M.}\ \bibnamefont {Newman}}\ and\ \bibinfo {author} {\bibfnamefont {D.~L.}\ \bibnamefont {Stein}},\ }\bibfield  {title} {\bibinfo {title} {Multiple states and thermodynamic limits in short-ranged ising spin-glass models},\ }\href {https://doi.org/10.1103/PhysRevB.46.973} {\bibfield  {journal} {\bibinfo  {journal} {Phys. Rev. B}\ }\textbf {\bibinfo {volume} {46}},\ \bibinfo {pages} {973} (\bibinfo {year} {1992})}\BibitemShut {NoStop}%
\bibitem [{\citenamefont {Newman}\ and\ \citenamefont {Stein}(1996)}]{newman:96b}%
  \BibitemOpen
  \bibfield  {author} {\bibinfo {author} {\bibfnamefont {C.~M.}\ \bibnamefont {Newman}}\ and\ \bibinfo {author} {\bibfnamefont {D.~L.}\ \bibnamefont {Stein}},\ }\bibfield  {title} {\bibinfo {title} {Spatial inhomogeneity and thermodynamic chaos},\ }\href {https://doi.org/10.1103/PhysRevLett.76.4821} {\bibfield  {journal} {\bibinfo  {journal} {Phys. Rev. Lett.}\ }\textbf {\bibinfo {volume} {76}},\ \bibinfo {pages} {4821} (\bibinfo {year} {1996})}\BibitemShut {NoStop}%
\bibitem [{\citenamefont {Newman}\ and\ \citenamefont {Stein}(1997)}]{newman:97}%
  \BibitemOpen
  \bibfield  {author} {\bibinfo {author} {\bibfnamefont {C.~M.}\ \bibnamefont {Newman}}\ and\ \bibinfo {author} {\bibfnamefont {D.~L.}\ \bibnamefont {Stein}},\ }\bibfield  {title} {\bibinfo {title} {Metastate approach to thermodynamic chaos},\ }\href {https://doi.org/10.1103/PhysRevE.55.5194} {\bibfield  {journal} {\bibinfo  {journal} {Phys. Rev. E}\ }\textbf {\bibinfo {volume} {55}},\ \bibinfo {pages} {5194} (\bibinfo {year} {1997})}\BibitemShut {NoStop}%
\bibitem [{\citenamefont {Newman}\ and\ \citenamefont {Stein}(1998)}]{newman:98}%
  \BibitemOpen
  \bibfield  {author} {\bibinfo {author} {\bibfnamefont {C.~M.}\ \bibnamefont {Newman}}\ and\ \bibinfo {author} {\bibfnamefont {D.~L.}\ \bibnamefont {Stein}},\ }\bibfield  {title} {\bibinfo {title} {Simplicity of state and overlap structure in finite-volume realistic spin glasses},\ }\href {https://doi.org/10.1103/PhysRevE.57.1356} {\bibfield  {journal} {\bibinfo  {journal} {Phys. Rev. E}\ }\textbf {\bibinfo {volume} {57}},\ \bibinfo {pages} {1356} (\bibinfo {year} {1998})}\BibitemShut {NoStop}%
\bibitem [{Note1()}]{Note1}%
  \BibitemOpen
  \bibinfo {note} {In this paper, finite-dimensional spin glasses mean spin glasses at a dimensionality above the lower critical dimension and below the upper critical one. One can assume 5/2 \cite {maiorano:18, bernaschi:26b} as the lower critical dimension and 6 \cite {aguilar-janita:24b, aguilar-janita:25} as the upper one, although there is no general consensus~\cite {fisher:88, newman:25}.}\BibitemShut {Stop}%
\bibitem [{\citenamefont {Marinari}\ \emph {et~al.}(2000)\citenamefont {Marinari}, \citenamefont {Parisi}, \citenamefont {Ricci-Tersenghi}, \citenamefont {Ruiz-Lorenzo},\ and\ \citenamefont {Zuliani}}]{marinari:00}%
  \BibitemOpen
  \bibfield  {author} {\bibinfo {author} {\bibfnamefont {E.}~\bibnamefont {Marinari}}, \bibinfo {author} {\bibfnamefont {G.}~\bibnamefont {Parisi}}, \bibinfo {author} {\bibfnamefont {F.}~\bibnamefont {Ricci-Tersenghi}}, \bibinfo {author} {\bibfnamefont {J.~J.}\ \bibnamefont {Ruiz-Lorenzo}},\ and\ \bibinfo {author} {\bibfnamefont {F.}~\bibnamefont {Zuliani}},\ }\bibfield  {title} {\bibinfo {title} {Replica symmetry breaking in short-range spin glasses: Theoretical foundations and numerical evidences},\ }\href {https://doi.org/10.1023/A:1018607809852} {\bibfield  {journal} {\bibinfo  {journal} {J. Stat. Phys.}\ }\textbf {\bibinfo {volume} {98}},\ \bibinfo {pages} {973} (\bibinfo {year} {2000})}\BibitemShut {NoStop}%
\bibitem [{\citenamefont {{Martin-Mayor}}\ \emph {et~al.}(2023)\citenamefont {{Martin-Mayor}}, \citenamefont {{Ruiz-Lorenzo}}, \citenamefont {{Seoane}},\ and\ \citenamefont {{Young}}}]{martin-mayor:22}%
  \BibitemOpen
  \bibfield  {author} {\bibinfo {author} {\bibfnamefont {V.}~\bibnamefont {{Martin-Mayor}}}, \bibinfo {author} {\bibfnamefont {J.}~\bibnamefont {{Ruiz-Lorenzo}}}, \bibinfo {author} {\bibfnamefont {B.}~\bibnamefont {{Seoane}}},\ and\ \bibinfo {author} {\bibfnamefont {P.}~\bibnamefont {{Young}}},\ }\bibfield  {title} {\bibinfo {title} {Numerical simulations and replica symmetry breaking},\ }in\ \href {https://doi.org/10.1142/13341} {\emph {\bibinfo {booktitle} {Spin Glass Theory and Far Beyond Replica Symmetry Breaking after 40 Years}}}\ (\bibinfo  {publisher} {World Scientific},\ \bibinfo {address} {Singapur},\ \bibinfo {year} {2023})\BibitemShut {NoStop}%
\bibitem [{\citenamefont {Ba\~nos}\ \emph {et~al.}(2011)\citenamefont {Ba\~nos}, \citenamefont {Cruz}, \citenamefont {Fernandez}, \citenamefont {Gil-Narvion}, \citenamefont {Gordillo-Guerrero}, \citenamefont {Guidetti}, \citenamefont {I\~niguez}, \citenamefont {Maiorano}, \citenamefont {Mantovani}, \citenamefont {Marinari}, \citenamefont {Mart\'{i}n-Mayor}, \citenamefont {Monforte-Garcia}, \citenamefont {Mu\~noz Sudupe}, \citenamefont {Navarro}, \citenamefont {Parisi}, \citenamefont {Perez-Gaviro}, \citenamefont {Ricci-Tersenghi}, \citenamefont {Ruiz-Lorenzo}, \citenamefont {Schifano}, \citenamefont {Seoane}, \citenamefont {Taranc\'on}, \citenamefont {Tripiccione},\ and\ \citenamefont {Yllanes}}]{janus:11}%
  \BibitemOpen
  \bibfield  {author} {\bibinfo {author} {\bibfnamefont {R.~A.}\ \bibnamefont {Ba\~nos}}, \bibinfo {author} {\bibfnamefont {A.}~\bibnamefont {Cruz}}, \bibinfo {author} {\bibfnamefont {L.~A.}\ \bibnamefont {Fernandez}}, \bibinfo {author} {\bibfnamefont {J.~M.}\ \bibnamefont {Gil-Narvion}}, \bibinfo {author} {\bibfnamefont {A.}~\bibnamefont {Gordillo-Guerrero}}, \bibinfo {author} {\bibfnamefont {M.}~\bibnamefont {Guidetti}}, \bibinfo {author} {\bibfnamefont {D.}~\bibnamefont {I\~niguez}}, \bibinfo {author} {\bibfnamefont {A.}~\bibnamefont {Maiorano}}, \bibinfo {author} {\bibfnamefont {F.}~\bibnamefont {Mantovani}}, \bibinfo {author} {\bibfnamefont {E.}~\bibnamefont {Marinari}}, \bibinfo {author} {\bibfnamefont {V.}~\bibnamefont {Mart\'{i}n-Mayor}}, \bibinfo {author} {\bibfnamefont {J.}~\bibnamefont {Monforte-Garcia}}, \bibinfo {author} {\bibfnamefont {A.}~\bibnamefont {Mu\~noz Sudupe}}, \bibinfo {author} {\bibfnamefont {D.}~\bibnamefont {Navarro}}, \bibinfo {author} {\bibfnamefont {G.}~\bibnamefont {Parisi}},
  \bibinfo {author} {\bibfnamefont {S.}~\bibnamefont {Perez-Gaviro}}, \bibinfo {author} {\bibfnamefont {F.}~\bibnamefont {Ricci-Tersenghi}}, \bibinfo {author} {\bibfnamefont {J.~J.}\ \bibnamefont {Ruiz-Lorenzo}}, \bibinfo {author} {\bibfnamefont {S.~F.}\ \bibnamefont {Schifano}}, \bibinfo {author} {\bibfnamefont {B.}~\bibnamefont {Seoane}}, \bibinfo {author} {\bibfnamefont {A.}~\bibnamefont {Taranc\'on}}, \bibinfo {author} {\bibfnamefont {R.}~\bibnamefont {Tripiccione}},\ and\ \bibinfo {author} {\bibfnamefont {D.}~\bibnamefont {Yllanes}},\ }\bibfield  {title} {\bibinfo {title} {Sample-to-sample fluctuations of the overlap distributions in the three-dimensional edwards-anderson spin glass},\ }\href {https://doi.org/10.1103/PhysRevB.84.174209} {\bibfield  {journal} {\bibinfo  {journal} {Phys. Rev. B}\ }\textbf {\bibinfo {volume} {84}},\ \bibinfo {pages} {174209} (\bibinfo {year} {2011})},\ \Eprint {https://arxiv.org/abs/arXiv:1107.5772} {arXiv:1107.5772} \BibitemShut {NoStop}%
\bibitem [{\citenamefont {Ghirlanda}\ and\ \citenamefont {Guerra}(1998)}]{ghirlanda:98}%
  \BibitemOpen
  \bibfield  {author} {\bibinfo {author} {\bibfnamefont {S.}~\bibnamefont {Ghirlanda}}\ and\ \bibinfo {author} {\bibfnamefont {F.}~\bibnamefont {Guerra}},\ }\href {https://doi.org/10.1088/0305-4470/31/46/006} {\bibfield  {journal} {\bibinfo  {journal} {J. Phys. A: Math. Gen.}\ }\textbf {\bibinfo {volume} {31}},\ \bibinfo {pages} {9149} (\bibinfo {year} {1998})},\ \Eprint {https://arxiv.org/abs/arXiv:cond-mat/9807333} {arXiv:cond-mat/9807333} \BibitemShut {NoStop}%
\bibitem [{\citenamefont {Franz}\ \emph {et~al.}(1998)\citenamefont {Franz}, \citenamefont {M\'ezard}, \citenamefont {Parisi},\ and\ \citenamefont {Peliti}}]{franz:98}%
  \BibitemOpen
  \bibfield  {author} {\bibinfo {author} {\bibfnamefont {S.}~\bibnamefont {Franz}}, \bibinfo {author} {\bibfnamefont {M.}~\bibnamefont {M\'ezard}}, \bibinfo {author} {\bibfnamefont {G.}~\bibnamefont {Parisi}},\ and\ \bibinfo {author} {\bibfnamefont {L.}~\bibnamefont {Peliti}},\ }\bibfield  {title} {\bibinfo {title} {Measuring equilibrium properties in aging systems},\ }\href {https://doi.org/10.1103/PhysRevLett.81.1758} {\bibfield  {journal} {\bibinfo  {journal} {Phys. Rev. Lett.}\ }\textbf {\bibinfo {volume} {81}},\ \bibinfo {pages} {1758} (\bibinfo {year} {1998})}\BibitemShut {NoStop}%
\bibitem [{\citenamefont {Franz}\ and\ \citenamefont {Ricci-Tersenghi}(2000)}]{franz2000ultrametricity}%
  \BibitemOpen
  \bibfield  {author} {\bibinfo {author} {\bibfnamefont {S.}~\bibnamefont {Franz}}\ and\ \bibinfo {author} {\bibfnamefont {F.}~\bibnamefont {Ricci-Tersenghi}},\ }\bibfield  {title} {\bibinfo {title} {Ultrametricity in three-dimensional edwards-anderson spin glasses},\ }\href {https://doi.org/10.1103/physreve.61.1121} {\bibfield  {journal} {\bibinfo  {journal} {Physical Review E}\ }\textbf {\bibinfo {volume} {61}},\ \bibinfo {pages} {1121} (\bibinfo {year} {2000})}\BibitemShut {NoStop}%
\bibitem [{\citenamefont {Contucci}\ \emph {et~al.}(2007)\citenamefont {Contucci}, \citenamefont {Giardin{\`a}}, \citenamefont {Giberti}, \citenamefont {Parisi},\ and\ \citenamefont {Vernia}}]{contucci:07b}%
  \BibitemOpen
  \bibfield  {author} {\bibinfo {author} {\bibfnamefont {P.}~\bibnamefont {Contucci}}, \bibinfo {author} {\bibfnamefont {C.}~\bibnamefont {Giardin{\`a}}}, \bibinfo {author} {\bibfnamefont {C.}~\bibnamefont {Giberti}}, \bibinfo {author} {\bibfnamefont {G.}~\bibnamefont {Parisi}},\ and\ \bibinfo {author} {\bibfnamefont {C.}~\bibnamefont {Vernia}},\ }\bibfield  {title} {\bibinfo {title} {Ultrametricity in the {E}dwards-{A}nderson model},\ }\href {https://doi.org/10.1103/PhysRevLett.99.057206} {\bibfield  {journal} {\bibinfo  {journal} {Phys. Rev. Lett}\ }\textbf {\bibinfo {volume} {99}},\ \bibinfo {pages} {057206} (\bibinfo {year} {2007})},\ \Eprint {https://arxiv.org/abs/arXiv:cond-mat/0607376} {arXiv:cond-mat/0607376} \BibitemShut {NoStop}%
\bibitem [{\citenamefont {Aguilar-Janita}\ \emph {et~al.}(2024{\natexlab{a}})\citenamefont {Aguilar-Janita}, \citenamefont {Franz}, \citenamefont {Martin-Mayor}, \citenamefont {Moreno-Gordo}, \citenamefont {Parisi}, \citenamefont {Ricci-Tersenghi},\ and\ \citenamefont {Ruiz-Lorenzo}}]{aguilar-janita:24}%
  \BibitemOpen
  \bibfield  {author} {\bibinfo {author} {\bibfnamefont {M.}~\bibnamefont {Aguilar-Janita}}, \bibinfo {author} {\bibfnamefont {S.}~\bibnamefont {Franz}}, \bibinfo {author} {\bibfnamefont {V.}~\bibnamefont {Martin-Mayor}}, \bibinfo {author} {\bibfnamefont {J.}~\bibnamefont {Moreno-Gordo}}, \bibinfo {author} {\bibfnamefont {G.}~\bibnamefont {Parisi}}, \bibinfo {author} {\bibfnamefont {F.}~\bibnamefont {Ricci-Tersenghi}},\ and\ \bibinfo {author} {\bibfnamefont {J.~J.}\ \bibnamefont {Ruiz-Lorenzo}},\ }\bibfield  {title} {\bibinfo {title} {Small field chaos in spin glasses: Universal predictions from the ultrametric tree and comparison with numerical simulations},\ }\href {https://doi.org/10.1073/pnas.2404973121} {\bibfield  {journal} {\bibinfo  {journal} {Proceedings of the National Academy of Sciences}\ }\textbf {\bibinfo {volume} {121}},\ \bibinfo {pages} {e2404973121} (\bibinfo {year} {2024}{\natexlab{a}})}\BibitemShut {NoStop}%
\bibitem [{\citenamefont {Baity-Jesi}\ \emph {et~al.}(2013)\citenamefont {Baity-Jesi}, \citenamefont {Ba\~{n}os}, \citenamefont {Cruz}, \citenamefont {Fernandez}, \citenamefont {Gil-Narvion}, \citenamefont {Gordillo-Guerrero}, \citenamefont {Iniguez}, \citenamefont {Maiorano}, \citenamefont {Mantovani}, \citenamefont {Marinari}, \citenamefont {Mart\'{i}n-Mayor}, \citenamefont {Monforte-Garcia}, \citenamefont {Mu{\~n}oz~Sudupe}, \citenamefont {Navarro}, \citenamefont {Parisi}, \citenamefont {Perez-Gaviro}, \citenamefont {Pivanti}, \citenamefont {Ricci-Tersenghi}, \citenamefont {Ruiz-Lorenzo}, \citenamefont {Schifano}, \citenamefont {Seoane}, \citenamefont {Tarancon}, \citenamefont {Tripiccione},\ and\ \citenamefont {Yllanes}}]{janus:13}%
  \BibitemOpen
  \bibfield  {author} {\bibinfo {author} {\bibfnamefont {M.}~\bibnamefont {Baity-Jesi}}, \bibinfo {author} {\bibfnamefont {R.~A.}\ \bibnamefont {Ba\~{n}os}}, \bibinfo {author} {\bibfnamefont {A.}~\bibnamefont {Cruz}}, \bibinfo {author} {\bibfnamefont {L.~A.}\ \bibnamefont {Fernandez}}, \bibinfo {author} {\bibfnamefont {J.~M.}\ \bibnamefont {Gil-Narvion}}, \bibinfo {author} {\bibfnamefont {A.}~\bibnamefont {Gordillo-Guerrero}}, \bibinfo {author} {\bibfnamefont {D.}~\bibnamefont {Iniguez}}, \bibinfo {author} {\bibfnamefont {A.}~\bibnamefont {Maiorano}}, \bibinfo {author} {\bibfnamefont {F.}~\bibnamefont {Mantovani}}, \bibinfo {author} {\bibfnamefont {E.}~\bibnamefont {Marinari}}, \bibinfo {author} {\bibfnamefont {V.}~\bibnamefont {Mart\'{i}n-Mayor}}, \bibinfo {author} {\bibfnamefont {J.}~\bibnamefont {Monforte-Garcia}}, \bibinfo {author} {\bibfnamefont {A.}~\bibnamefont {Mu{\~n}oz~Sudupe}}, \bibinfo {author} {\bibfnamefont {D.}~\bibnamefont {Navarro}}, \bibinfo {author} {\bibfnamefont {G.}~\bibnamefont
  {Parisi}}, \bibinfo {author} {\bibfnamefont {S.}~\bibnamefont {Perez-Gaviro}}, \bibinfo {author} {\bibfnamefont {M.}~\bibnamefont {Pivanti}}, \bibinfo {author} {\bibfnamefont {F.}~\bibnamefont {Ricci-Tersenghi}}, \bibinfo {author} {\bibfnamefont {J.~J.}\ \bibnamefont {Ruiz-Lorenzo}}, \bibinfo {author} {\bibfnamefont {S.~F.}\ \bibnamefont {Schifano}}, \bibinfo {author} {\bibfnamefont {B.}~\bibnamefont {Seoane}}, \bibinfo {author} {\bibfnamefont {A.}~\bibnamefont {Tarancon}}, \bibinfo {author} {\bibfnamefont {R.}~\bibnamefont {Tripiccione}},\ and\ \bibinfo {author} {\bibfnamefont {D.}~\bibnamefont {Yllanes}} (\bibinfo {collaboration} {Janus Collaboration}),\ }\bibfield  {title} {\bibinfo {title} {Critical parameters of the three-dimensional {I}sing spin glass},\ }\href {https://doi.org/10.1103/PhysRevB.88.224416} {\bibfield  {journal} {\bibinfo  {journal} {Phys. Rev. B}\ }\textbf {\bibinfo {volume} {88}},\ \bibinfo {pages} {224416} (\bibinfo {year} {{2013}})}\BibitemShut {NoStop}%
\bibitem [{\citenamefont {Billoire}\ \emph {et~al.}(2018)\citenamefont {Billoire}, \citenamefont {Fernandez}, \citenamefont {Maiorano}, \citenamefont {Marinari}, \citenamefont {Martin-Mayor}, \citenamefont {Moreno-Gordo}, \citenamefont {Parisi}, \citenamefont {Ricci-Tersenghi},\ and\ \citenamefont {Ruiz-Lorenzo}}]{billoire:18}%
  \BibitemOpen
  \bibfield  {author} {\bibinfo {author} {\bibfnamefont {A.}~\bibnamefont {Billoire}}, \bibinfo {author} {\bibfnamefont {L.~A.}\ \bibnamefont {Fernandez}}, \bibinfo {author} {\bibfnamefont {A.}~\bibnamefont {Maiorano}}, \bibinfo {author} {\bibfnamefont {E.}~\bibnamefont {Marinari}}, \bibinfo {author} {\bibfnamefont {V.}~\bibnamefont {Martin-Mayor}}, \bibinfo {author} {\bibfnamefont {J.}~\bibnamefont {Moreno-Gordo}}, \bibinfo {author} {\bibfnamefont {G.}~\bibnamefont {Parisi}}, \bibinfo {author} {\bibfnamefont {F.}~\bibnamefont {Ricci-Tersenghi}},\ and\ \bibinfo {author} {\bibfnamefont {J.~J.}\ \bibnamefont {Ruiz-Lorenzo}},\ }\bibfield  {title} {\bibinfo {title} {Dynamic variational study of chaos: spin glasses in three dimensions},\ }\href {https://doi.org/10.1088/1742-5468/aaa387} {\bibfield  {journal} {\bibinfo  {journal} {J. Stat. Mech.}\ }\textbf {\bibinfo {volume} {2018}},\ \bibinfo {pages} {033302} (\bibinfo {year} {2018})}\BibitemShut {NoStop}%
\bibitem [{Note2()}]{Note2}%
  \BibitemOpen
  \bibinfo {note} {Note that we do not introduce any binning in $q$ for the computation of $P(q)$; instead, we consider all the $2N+1$ possible values of $q$.}\BibitemShut {Stop}%
\bibitem [{Note3()}]{Note3}%
  \BibitemOpen
  \bibinfo {note} {To generate a clear comparison, we have divided the numerical $P^L(q);0$ into a given number of equal-probability intervals.}\BibitemShut {Stop}%
\bibitem [{Note4()}]{Note4}%
  \BibitemOpen
  \bibinfo {note} {To minimize systematic errors, we used the same samples to compute the observables in the numerator and the denominator.}\BibitemShut {Stop}%
\bibitem [{Note5()}]{Note5}%
  \BibitemOpen
  \bibinfo {note} {This functional form is a Padé approximant for the odd function $q(x)$. Note that $q(x)$ is an odd function because also $x(q)$ is an odd function, see Eq.~\protect \eqref {eq:x-q-def}.}\BibitemShut {Stop}%
\bibitem [{\citenamefont {Maiorano}\ and\ \citenamefont {Parisi}(2018)}]{maiorano:18}%
  \BibitemOpen
  \bibfield  {author} {\bibinfo {author} {\bibfnamefont {A.}~\bibnamefont {Maiorano}}\ and\ \bibinfo {author} {\bibfnamefont {G.}~\bibnamefont {Parisi}},\ }\bibfield  {title} {\bibinfo {title} {Support for the value 5/2 for the spin glass lower critical dimension at zero magnetic field},\ }\href {https://doi.org/10.1073/pnas.1720832115} {\bibfield  {journal} {\bibinfo  {journal} {Proceedings of the National Academy of Sciences}\ }\textbf {\bibinfo {volume} {115}},\ \bibinfo {pages} {5129} (\bibinfo {year} {2018})},\ \Eprint {https://arxiv.org/abs/https://www.pnas.org/content/115/20/5129.full.pdf} {https://www.pnas.org/content/115/20/5129.full.pdf} \BibitemShut {NoStop}%
\bibitem [{\citenamefont {Bernaschi}\ \emph {et~al.}(2026)\citenamefont {Bernaschi}, \citenamefont {Fernandez}, \citenamefont {González-Adalid~Pemartín}, \citenamefont {Martin-Mayor}, \citenamefont {Parisi},\ and\ \citenamefont {Ricci-Tersenghi}}]{bernaschi:26b}%
  \BibitemOpen
  \bibfield  {author} {\bibinfo {author} {\bibfnamefont {M.}~\bibnamefont {Bernaschi}}, \bibinfo {author} {\bibfnamefont {L.~A.}\ \bibnamefont {Fernandez}}, \bibinfo {author} {\bibfnamefont {I.}~\bibnamefont {González-Adalid~Pemartín}}, \bibinfo {author} {\bibfnamefont {V.}~\bibnamefont {Martin-Mayor}}, \bibinfo {author} {\bibfnamefont {G.}~\bibnamefont {Parisi}},\ and\ \bibinfo {author} {\bibfnamefont {F.}~\bibnamefont {Ricci-Tersenghi}},\ }\href@noop {} {\bibinfo {title} {Low energy excitations in a long prism geometry: computing the lower critical dimension of the ising spin glass}} (\bibinfo {year} {2026}),\ \Eprint {https://arxiv.org/abs/2601.07926} {arXiv:2601.07926 [cond-mat.stat-mech]} \BibitemShut {NoStop}%
\bibitem [{\citenamefont {Aguilar-Janita}\ \emph {et~al.}(2024{\natexlab{b}})\citenamefont {Aguilar-Janita}, \citenamefont {Martin-Mayor}, \citenamefont {Moreno-Gordo},\ and\ \citenamefont {Ruiz-Lorenzo}}]{aguilar-janita:24b}%
  \BibitemOpen
  \bibfield  {author} {\bibinfo {author} {\bibfnamefont {M.}~\bibnamefont {Aguilar-Janita}}, \bibinfo {author} {\bibfnamefont {V.}~\bibnamefont {Martin-Mayor}}, \bibinfo {author} {\bibfnamefont {J.}~\bibnamefont {Moreno-Gordo}},\ and\ \bibinfo {author} {\bibfnamefont {J.~J.}\ \bibnamefont {Ruiz-Lorenzo}},\ }\bibfield  {title} {\bibinfo {title} {Evidence of a second-order phase transition in the six-dimensional ising spin glass in a field},\ }\href {https://doi.org/10.1103/PhysRevE.109.055302} {\bibfield  {journal} {\bibinfo  {journal} {Phys. Rev. E}\ }\textbf {\bibinfo {volume} {109}},\ \bibinfo {pages} {055302} (\bibinfo {year} {2024}{\natexlab{b}})}\BibitemShut {NoStop}%
\bibitem [{\citenamefont {Aguilar-Janita}\ \emph {et~al.}(2025)\citenamefont {Aguilar-Janita}, \citenamefont {Martín-Mayor}, \citenamefont {Moreno-Gordo},\ and\ \citenamefont {Ruiz-Lorenzo}}]{aguilar-janita:25}%
  \BibitemOpen
  \bibfield  {author} {\bibinfo {author} {\bibfnamefont {M.}~\bibnamefont {Aguilar-Janita}}, \bibinfo {author} {\bibfnamefont {V.}~\bibnamefont {Martín-Mayor}}, \bibinfo {author} {\bibfnamefont {J.}~\bibnamefont {Moreno-Gordo}},\ and\ \bibinfo {author} {\bibfnamefont {J.~J.}\ \bibnamefont {Ruiz-Lorenzo}},\ }\bibfield  {title} {\bibinfo {title} {Evidence of de almeida–thouless line below six dimensions},\ }\href {https://doi.org/10.1088/1742-5468/ae1a4a} {\bibfield  {journal} {\bibinfo  {journal} {Journal of Statistical Mechanics: Theory and Experiment}\ }\textbf {\bibinfo {volume} {2025}},\ \bibinfo {pages} {113301} (\bibinfo {year} {2025})}\BibitemShut {NoStop}%
\bibitem [{\citenamefont {Newman}\ and\ \citenamefont {Stein}(2025)}]{newman:25}%
  \BibitemOpen
  \bibfield  {author} {\bibinfo {author} {\bibfnamefont {C.~M.}\ \bibnamefont {Newman}}\ and\ \bibinfo {author} {\bibfnamefont {D.~L.}\ \bibnamefont {Stein}},\ }\bibfield  {title} {\bibinfo {title} {{G}round state excitations and energy fluctuations in short-range spin glasses},\ }\Eprint {https://arxiv.org/abs/arXiv:2510.27507} {arXiv:2510.27507}  (\bibinfo {year} {2025}),\ \bibinfo {note} {unpublished}\BibitemShut {NoStop}%
\bibitem [{\citenamefont {Newman}\ and\ \citenamefont {Barkema}(1999)}]{newman:99}%
  \BibitemOpen
  \bibfield  {author} {\bibinfo {author} {\bibfnamefont {M.~E.~J.}\ \bibnamefont {Newman}}\ and\ \bibinfo {author} {\bibfnamefont {G.~T.}\ \bibnamefont {Barkema}},\ }\href@noop {} {\emph {\bibinfo {title} {{M}onte {C}arlo Methods in Statistical Physics}}}\ (\bibinfo  {publisher} {Clarendon Press},\ \bibinfo {address} {Oxford},\ \bibinfo {year} {1999})\BibitemShut {NoStop}%
\bibitem [{\citenamefont {Hukushima}\ and\ \citenamefont {Nemoto}(1996)}]{hukushima:96}%
  \BibitemOpen
  \bibfield  {author} {\bibinfo {author} {\bibfnamefont {K.}~\bibnamefont {Hukushima}}\ and\ \bibinfo {author} {\bibfnamefont {K.}~\bibnamefont {Nemoto}},\ }\bibfield  {title} {\bibinfo {title} {{E}xchange {M}onte {C}arlo method and application to spin glass simulations},\ }\href {https://doi.org/10.1143/JPSJ.65.1604} {\bibfield  {journal} {\bibinfo  {journal} {J. Phys. Soc. Japan}\ }\textbf {\bibinfo {volume} {65}},\ \bibinfo {pages} {1604} (\bibinfo {year} {1996})}\BibitemShut {NoStop}%
\bibitem [{\citenamefont {Marinari}\ \emph {et~al.}(1998)\citenamefont {Marinari}, \citenamefont {Parisi},\ and\ \citenamefont {Rossetti}}]{marinari:98}%
  \BibitemOpen
  \bibfield  {author} {\bibinfo {author} {\bibfnamefont {E.}~\bibnamefont {Marinari}}, \bibinfo {author} {\bibfnamefont {G.}~\bibnamefont {Parisi}},\ and\ \bibinfo {author} {\bibfnamefont {D.}~\bibnamefont {Rossetti}},\ }\href@noop {} {\bibfield  {journal} {\bibinfo  {journal} {Eur. Phys. J. B}\ }\textbf {\bibinfo {volume} {2}},\ \bibinfo {pages} {495} (\bibinfo {year} {1998})}\BibitemShut {NoStop}%
\bibitem [{\citenamefont {Young}(2012)}]{young:12}%
  \BibitemOpen
  \bibfield  {author} {\bibinfo {author} {\bibfnamefont {A.~P.}\ \bibnamefont {Young}},\ }\href@noop {} {\bibinfo {title} {Everything you wanted to know about data analysis and fitting but were afraid to ask}},\ \bibinfo {howpublished} {School on "Efficient Algorithms in Computational Physics"},\ \bibinfo {address} {Bad Honnef} (\bibinfo {year} {2012}),\ \Eprint {https://arxiv.org/abs/arXiv:1210.3781v3} {arXiv:1210.3781v3} \BibitemShut {NoStop}%
\bibitem [{\citenamefont {Read}(2014)}]{read:14}%
  \BibitemOpen
  \bibfield  {author} {\bibinfo {author} {\bibfnamefont {N.}~\bibnamefont {Read}},\ }\bibfield  {title} {\bibinfo {title} {Short-range {I}sing spin glasses: The metastate interpretation of replica symmetry breaking},\ }\href {https://doi.org/10.1103/PhysRevE.90.032142} {\bibfield  {journal} {\bibinfo  {journal} {Phys. Rev. E}\ }\textbf {\bibinfo {volume} {90}},\ \bibinfo {pages} {032142} (\bibinfo {year} {2014})}\BibitemShut {NoStop}%
\bibitem [{Note6()}]{Note6}%
  \BibitemOpen
  \bibinfo {note} {Because the sample-resolved data were lost during the final stages of this work, the error bars for $L=12$ and $16$ in Figs.~\ref {fig:rstraight_multifield} and \ref {fig:rchaos_multifield} could not be estimated using the jackknife procedure employed for the $L=8$ data and for all other error bars in the paper. Instead, we obtained them by Gaussian resampling, propagating the uncertainty in the mean $P(q)$. Because this procedure systematically underestimates the uncertainties, we rescaled the resulting errors using correction factors determined, for each value of $h$, by comparing the propagated and jackknife errors at $L=8$. For $h=2,4,6,$ and $8$, respectively, these factors are $2.51$, $3.36$, $3.69$, and $3.82$ for the straight protocol, and $3.10$, $5.65$, $7.11$, and $8.03$ for the chaos protocol. Consequently, the error bars for $L=12$ and $16$ in these figures should be regarded only as order-of-magnitude estimates.}\BibitemShut {Stop}%
\end{thebibliography}%

\appendix

\section{Details of the numerical simulations}
\label{app:num}

All the simulations of this work have been performed in a cluster with eight different nodes Intel\registeredmark Xeon\registeredmark Gold 6240 CPUs (18 cores per socket at 2.60 GHz), giving 36 cores per node and 288 cores in total.

We take advantage of the multispin coding to simulate our spin-glass system. We first use MUSA (MUlti SAmple multispin coding)~\cite{newman:99}, which makes intensive use of the 128-bit registers of modern CPUs. We have simulated
128 different systems by using AVX instructions to perform atomic bitwise operations.

Moreover, we use the Parallel Tempering method~\cite{hukushima:96,marinari:98} to reach thermal equilibrium in our system. Therefore, we simulate different copies of each sample at different temperatures. Table \ref{tab:sim_parameters} lists the simulated parameters. 

For each sample, we simulate 211353600 Metropolis sweeps, and for each 20 Metropolis sweeps we perform a Parallel Tempering attempt.

\begin{table}[h]
\centering
\begin{tabular}{c c c c c c} \hline \hline
$L$ & ~~\#Samples~~~~& Magnetic fields ($h$) &\#Temp. & ~$T_\mathrm{min}$ & ~$T_\mathrm{max}$ \\ 
\hline 
8 & 20 $\times$ 128 & \{0,1,2,3,4,6,8,10\} &10  &  0.805  & 1.432\\ 
12 & 20 $\times$ 128 & \{0,1,2,3,4,6,8,10\} &14  &  0.805  & 1.432\\ 
16 & 20 $\times$ 128 & \{0,1,2,3,4,6,8,10\} &24  &  0.805  &  1.432\\ 
\hline \hline
\end{tabular}
\caption{Some parameters of our simulation. $L$ refers to the linear size of the hypercube. The number of spins is $N=L^3$. For all the sizes, we simulate 20 $\times$ 128 samples. The $\times$ 128 specification emphasizes that each super-sample contains 128 samples. We simulate several external magnetic fields $h$ for each sample. For each system size, we select the appropriate number of temperatures that allow a correct exchange rate, ensuring the random walk in temperatures is sufficiently ergodic. Finally, the last two columns show the lower and upper bounds of the temperature interval used in parallel tempering.}
\label{tab:sim_parameters}
\end{table}

To ensure proper thermalization, we study each sample individually by following the procedure introduced in Ref.~\cite{billoire:18}. The thermalization starts by simulating a super sample for a large number of Metropolis sweeps determined by preliminary tests.

We monitor the temperature random walk during the simulation, then compute the integrated autocorrelation time from it. The integrated autocorrelation time $\tau_{\text{int},f}$ depends on the specific temperature function $f(T)$ used; see Ref.~\cite{billoire:18} for details. Since the exponential time $\tau_{\mathrm{exp}}$ is an upper bound of the integrated autocorrelation time, we follow a variational method and take the maximum of all the integrated autocorrelation times, making the assumption that $\max(\tau_{\text{int},f}) \sim \tau_\text{exp}$.

This estimate of $\tau_{\mathrm{exp}}$ lets us check whether a given sample is thermalized. We established a demanding thermalization threshold $N_\text{sim} > 30\, \tau_\text{exp}$.

In this simulation, we did not need to extract individual samples from the supersamples to complete thermalization~\cite {aguilar-janita:25}. 

We have used the jackknife method~\cite{young:12} to estimate the errors of the different numerical observables computed in this work.

\section{RSB and ultrametricity}
\label{app:ultra1}
One of the most remarkable predictions of RSB theory is that spin glasses at equilibrium have an infinite number of pure states, labeled by $\alpha$, which are organized in an ultrametric fashion. Those pure states are defined by the clustering property 
and have random statistical weights $w_\alpha$, such that the Gibbs measure can be decomposed as~\cite{mezard:87,read:14}
\begin{equation}
\langle (\cdots) \rangle    = \sum_\alpha w_\alpha \langle (\cdots) \rangle_{\alpha} \;. 
\end{equation} 

\begin{figure}[htpb]
\centering
\resizebox{0.48\textwidth}{!}{%
\begin{tikzpicture}[
    x=1cm, y=1cm,
    tree/.style={line width=1.6pt, line cap=round, line join=round},
    axis/.style={line width=1.8pt, -{Latex[length=4mm,width=3mm]}},
    dashedlevel/.style={line width=0.8pt, dash pattern=on 5pt off 5pt},
    lab/.style={font=\Large\bfseries},
    staten/.style={font=\Large\bfseries}
]

\def\qEA{0}
\def\qtwo{1.2}
\def\qone{2.4}
\def\qzero{4.8}

\coordinate (s1) at (1.0,\qEA);
\coordinate (s2) at (2.0,\qEA);
\coordinate (s3) at (3.0,\qEA);
\coordinate (s4) at (3.7,\qEA);
\coordinate (s5) at (4.7,\qEA);
\coordinate (s6) at (5.5,\qEA);
\coordinate (s7) at (6.8,\qEA);

\coordinate (a23)  at (2.0,\qtwo);
\coordinate (a45)  at (4.2,\qtwo);
\coordinate (a67)  at (6.1,\qtwo);
\coordinate (a123) at (4.2,\qone);
\coordinate (aTop) at (6.8,\qzero);
\coordinate (rTop) at (9.4,\qone);

\draw[axis] (0,-0.9) -- (0,5.45);

\draw[axis] (-0.45,-0.9) -- (11.3,-0.9);


\draw[dashedlevel] (0,\qzero) -- (6.8,\qzero);
\draw[dashedlevel] (0,\qone)  -- (4.2,\qone);
\draw[dashedlevel] (0,\qtwo)  -- (6.1,\qtwo);
\draw[dashedlevel] (0,\qEA)   -- (6.8,\qEA);

\draw[tree] (s1) -- (a23);
\draw[tree] (s2) -- (a23);
\draw[tree] (s3) -- (a23);

\draw[tree] (s4) -- (a45);
\draw[tree] (s5) -- (a45);

\draw[tree] (s6) -- (a67);
\draw[tree] (s7) -- (a67);

\draw[tree] (a23) -- (a123);
\draw[tree] (a45) -- (a123);
\draw[tree] (a67) -- (a123);

\draw[tree] (a123) -- (aTop);
\draw[tree] (aTop) -- (rTop);

\node[lab, anchor=east] at (-0.35,\qzero) {$q_0$};
\node[lab, anchor=east] at (-0.35,\qone)  {$q_1$};
\node[lab, anchor=east] at (-0.35,\qtwo)  {$q_2$};
\node[lab, anchor=east] at (-0.35,\qEA)   {$q_{\rm EA}$};

\node[staten, below=2pt] at (s1) {$\alpha$};
\node[staten, below=2pt] at (s2) {$\beta$};
\node[staten, below=2pt] at (s3) {$\gamma$};
\node[staten, below=2pt] at (s4) {$\delta$};
\node[staten, below=2pt] at (s5) {$\lambda$};
\node[staten, below=2pt] at (s6) {$\mu$};
\node[staten, below=2pt] at (s7) {$\nu$};

\node[font=\Large\itshape, anchor=north] at (9.0,-1.15) {pure states};
\node[font=\Large] at (9.5,1.75) {$\cdots$};

\end{tikzpicture}
}
\caption{Schematic representation of the hierarchical organization of pure states.}
\label{fig:states-tree}
\end{figure}
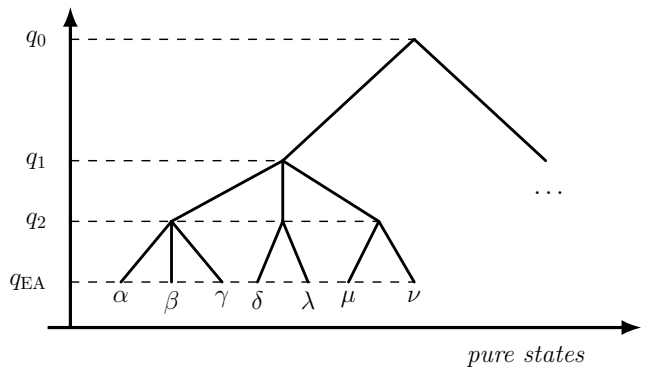

The ultrametric property means that, if one defines the distance between two pure states $\alpha$ and $\beta$
\begin{equation}
    d_{\alpha \beta}  = \frac{1}{2}(q_{\mathrm{EA}}-q_{\alpha\beta})
\end{equation}
it satisfies the condition 
\begin{equation}
    d_{\alpha\beta} \leq \max(d_{\alpha\gamma}, d_{\gamma\beta})\;, 
\end{equation}
where $q_{\alpha\beta}$ is defined as 
\begin{equation}
    q_{\alpha\beta}= \frac{1}{N} \sum_i \langle\sigma_i\rangle_{\alpha} \langle\sigma_i\rangle_{\beta} \;. 
\end{equation}
This ultrametric property means that in spin glasses described by RSB, pure states are organized in a (family) tree structure, where the leaves of the tree are the different pure states, and the overlap between them are given by the value associated with their nearest common ancestor; see Fig.~\ref{fig:states-tree}. It is important to note that a tree like the one represented in Fig.~\ref{fig:states-tree} describes the organization of pure states in a $k-$steps RSB scheme, but the solution for MF spin glasses is given by full-RSB, in which the tree has an infinite number of leaves and infinite branching points, meaning that the overlaps can take any value in the interval $q_{\alpha\beta}\in[-q_{\mathrm{EA}}, q_{\mathrm{EA}}] \subset \mathbb{R}$. 

In terms of pure states, the probability distribution of the overlaps $P(q)$ can be computed as 
\begin{equation}
    P_J(q)= \sum_{\alpha\beta} w_{\alpha}w_{\beta} \delta(q-q_{\alpha\beta})\;, \quad P(q)= \overline{P_J(q)}\;.
\end{equation}
This enables us to compute $P(q)$ once a family of ultrametric trees with the right statistical properties is constructed. 

A key property, central for our work, is the fact that, under a vanishing magnetic field $H=h/\sqrt{N}$, pure states keep their identity, but their relative weights are modified as 
\begin{equation}
    w_\alpha \to w_\alpha(h) \propto w_\alpha \exp(\beta h m_\alpha)\;,
\end{equation}
where $m_\alpha$ is the magnetization in the pure state $\alpha$ defined as
\begin{equation}
    m_\alpha = \frac{1}{\sqrt{N}}\sum_{i=1}^{N} \langle \sigma_i \rangle _\alpha \;,
\end{equation}
which in the absence of a magnetic field is of ${\cal O}(1)$.

 Another important feature is that, at $H=0$, the system is symmetric under spin reversal $\sigma \to -\sigma$. This implies that, for any sate $\alpha$, there is a symmetric state $\alpha^{\prime}$ with opposite magnetization $m_\alpha=-m_{\alpha^{\prime}}$ and same weight $w_{\alpha}=w_{\alpha ^{\prime}}$. Hence, we can use only one label for each pair of pure states such that $q_{\alpha\beta}>0 \; \forall (\alpha, \beta)$.

All of the above enables one to compute the probability distribution of the overlaps for a system in the presence of an external vanishing magnetic field $P(|q|,h)$, as well as the chaotic probability distribution of overlaps, defined in a system where one replica is at $H=0$ and the other one at $H>0$, $P^C(|q|,h)$. These observables can be computed as 
\begin{equation}\label{eq:app_Pc}
    P^C_J(|q|;h)=\sum_{\alpha\beta}w_\alpha w_\beta(h) \delta(q-q_{\alpha\beta})\;,
\end{equation}
with 
\begin{equation}
    w_\alpha(h)= \frac{w_\alpha \cosh(\beta h m_\alpha)}{\sum_\gamma w_\gamma \cosh(\beta h m_\gamma)}\;.
\end{equation}
Similarly, 
\begin{equation}\label{eq:app_Ph}
    P_J(\pm|q|;h) = \sum_{\alpha\beta} w_\alpha(h) w_\beta(h) g_{\pm}(h, m_\alpha,m_\beta)\delta(q-q_{\alpha\beta})\;,
\end{equation}
with 
\begin{equation}
    g_{\pm}(h,m_\alpha,m_\beta) = \frac{1\pm \tanh(\beta h m_\alpha)\tanh(\beta h m_\beta)}{2}\;.
\end{equation}

\section{Generating ultrametric trees}
\label{app:ultra2}
In order to measure $P(|q|;h)$ and $P^C(|q|;h)$ through equations \eqref{eq:app_Pc} and \eqref{eq:app_Ph} one needs to generate a set of ultrametric trees with the right statistical properties. For each of these trees, one must specify
\begin{itemize}
    \item The weights of the leaves $w_\alpha$.
    \item The overlaps between states $q_{\alpha\beta}$.
    \item The magnetizations of each pure state $m_\alpha$, which are random variables which follow a Gaussian distribution with zero mean and covariance $\overline{m_\alpha m_\beta}=q_{\alpha \beta}$. 
\end{itemize}
Although the tree has infinitely many states in principle, we neglect states with weight smaller than $\epsilon$, so the pruned tree has a finite number $M$ of leaves. 

The first step is to associate weights to each leaf. Since the weight can be written in terms of the free energies of pure states $f_\alpha$, and those follow an exponential distribution, it is easy to show that the unnormalized weights of pure states $u_\alpha=e^{-\beta f_\alpha}$ are distributed as 
\begin{equation}
    P(u) = \theta(u-1)x_{\mathrm{EA}}u^{-1-x_{\mathrm{EA}}}\;,
\end{equation}
and then $w_\alpha = u_{\alpha}/\sum_\gamma u_\gamma$. 

The structure of the tree is built following a Markovian process governed by the Bolthausen-Snitzman coalescence process~\cite{bolthausen:98} and is implemented as follows: 
\begin{enumerate}
    \item Start with $M$ leaves at $t=0$, and repeatedly coalesce $k$ nodes. The probability of coalescing $k$ nodes given that at the current level one has $b$ nodes is given by
    \begin{equation}
        P(k|b)=\frac{b}{(b-1)(k-1)k}\;.
    \end{equation}
    \item After each update, the time is increased $t\to t+\Delta t$ where the times increments $\Delta t$ follows an exponential distribution with mean $1/(b-1)$.
    \item Repeat the process until $b=1$.
\end{enumerate}
Once the tree structure is generated, the final step is to assign an overlap value $q$ to each node in the tree. The overlap is $q(x_{\mathrm{EA}}\exp(-t))$, where $t$ is the time label corresponding to the node. 

Finally, to obtain disordered-average quantities, one must average over different realizations of the ultrametric trees~\cite{aguilar-janita:24}.

\section{Checking $P(q)$} \label{app:checkPq}
We aim to perform a consistency check of our numerical results for three-dimensional spin glasses. As explained in the main text, we model the $P(q)$ in the thermodynamic limit as

\begin{equation}
\begin{split}
P(q) = &  P(0)(1 + \alpha q^2 + \gamma q^4) \,\Theta(\qeainfty - q) \\
& + \big[1 - x(\qeainfty)\big]\, \delta(q - \qeainfty) \, ,
\end{split}
\textbf{}
\end{equation}
where $\alpha$ and $\gamma$ are free coefficients fitted in the vicinity of $q=0$ and  $\qeainfty$ is the Edwards-Anderson order parameter in the thermodynamic limit.

We now proceed to perform a consistency check by effectively working backward: starting from the modeled curve in the thermodynamic limit, we follow the method introduced in Ref.~\cite{janus:11} to obtain the finite-size probability density function $P^L(q)$.

We introduce the variable change $q = \tanh(u)$ and fit the numerical $P^L(q)$ curve to the following distribution:
\begin{equation}
\begin{split}
\mathcal{P}(u) & = \big[1 - x(\qeainfty)\big] \Big(\frac{G^{(k)}(u - \ueainfty) + G^{(k)}(h + \ueainfty)}{2} \Big)\\
& + P(0) \int_{-\ueainfty}^{\ueainfty} \big(1 + \alpha z^2 + \gamma z^4\big) G^{(k)}(u - z) \dd z\,,
\end{split}
\end{equation}
where the convolution kernel is given by:
\begin{equation}
G^{(k)}(u - u') = C \exp\left( - \left| \frac{u - u'}{\Delta} \right|^k \right)\,,
\end{equation}
with $C$ and $\Delta$ being constants. The exponent $k$ was left free and we have obtained values near to $2.5$ as found by Ref. ~\cite{janus:11}. The parameters $\alpha$ and $\gamma$ are also treated as free fitting parameters, and the cumulative probability is evaluated analytically as:
\begin{equation}
x(\qeainfty) = 2 \tilde{P}(0) \Big( \ueainfty + \alpha \frac{{(\ueainfty})^3}{3} + \gamma \frac{{(\ueainfty)}^5}{5} \Big)\,.
\end{equation}

\begin{figure}[h]
    \centering
    \includegraphics[width=0.99\linewidth]{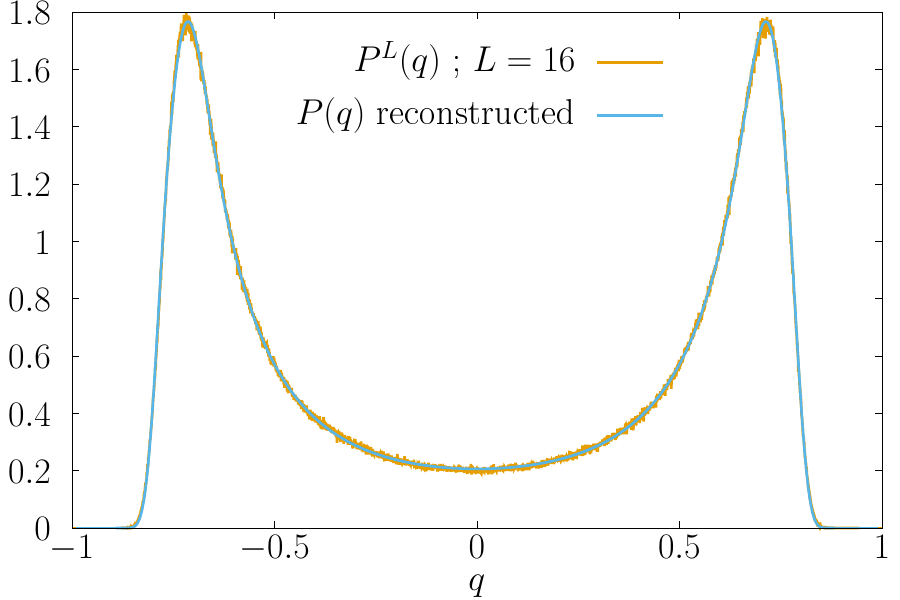}
    \caption{The original and reconstructed $P(q)$ for $L=16$.}
    \label{fig:reconstructed}
\end{figure}

Tab.~\ref{tab:ajuste_pq} shows the fitting results. The fit improves for larger system sizes but always provides a very good description of the numerical data, as shown in Fig.~\ref{fig:reconstructed}.

The reader may notice a discrepancy between $\qea$ in Table~\ref{tab:ajuste_pq_cuadratica} and $\qeainfty$ in Table~\ref{tab:ajuste_pq}. This difference stems from the physical meaning of each parameter and the role of finite-size effects.

\begin{table}[h]
    \centering
    \setlength{\tabcolsep}{3pt}
    \begin{tabular}{l c c c c c c c}
        \hline \hline
        $L$ & $\qeainfty$ & $\alpha$ & $\gamma$ & $\Delta$ & $k$ & $\chi^2/\text{d.o.f.}$ & $p$-value \\
        \hline
        8  &  0.831 & 0.896 & 1.656 & 0.255 & 2.00 & 353.98/497 & 0.999 \\
        12 & 0.785 & 1.552 & 3.452 & 0.182 & 2.28 &1414.09/1575 & 0.985 \\
        16 &  0.759 & 2.379 & 3.608 & 0.139 & 2.50 & 3006.07/3558 & 0.999 \\
        \hline \hline
    \end{tabular}
    \caption{Fitting parameters of the convoluted theoretical distribution to the numerical finite-size $P(q)$ for different system sizes $L$.}
    \label{tab:ajuste_pq}
\end{table}

The value $\qea$ from Table~\ref{tab:ajuste_pq_cuadratica} acts as an \textit{effective} parameter. In finite-size systems, the Dirac delta functions are broadened into bell-shaped curves. Because the continuous part of $P(q)$ grows as it approaches the Edwards-Anderson parameter, estimating $\qea$ by matching the first moment $\langle q \rangle$ inherently absorbs this finite-size shift, yielding a systematically lower value.

Conversely, the convolution method explicitly disentangles the finite-size broadening from the actual location of the peak. Thus, $\qeainfty$ represents the ``bare'' or asymptotic Edwards-Anderson order parameter in the thermodynamic limit and, therefore, is naturally larger than the effective $\qea$. 

The fact that we can accurately reconstruct the finite-size numerical density $P^L(q)$ by applying a size-dependent convolution kernel to the asymptotic theoretical distribution confirms the consistency of our data.

\section{Finite Size effects and higher fields ($h$)} 
\label{app:hf}

Here we examine how the comparison between the numerical overlap distributions and the ultrametric-tree predictions changes with the field and with the system size. We consider $h=2,4,6,$ and 8 for $L=8$ and 16, both for two replicas at the same field, $P^L(q;h)$ and for replicas at zero and nonzero field $P_\text{C}^L(q;h)$]. In every panel, the dashed gray curve is the numerical zero-field distribution $P^L(q)$, the solid colored curve is the theoretical prediction, and the points with vertical error bars are the numerical data at nonzero field.

\begin{figure*}
    \centering
    \includegraphics[width=0.87\textwidth]{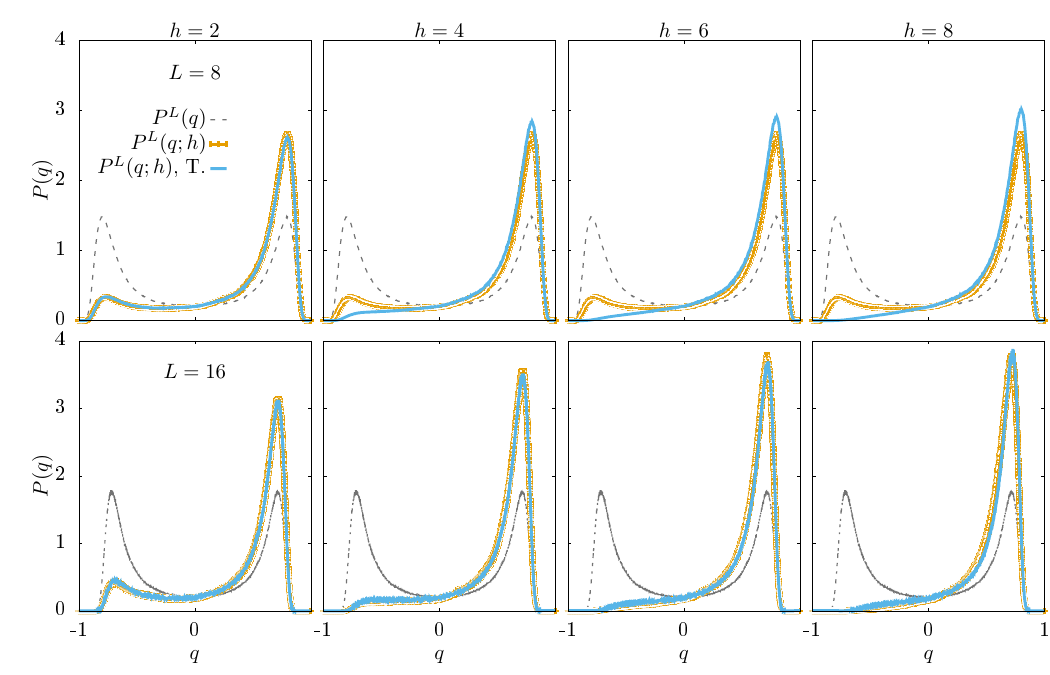}
    \caption{Comparison of $P^L(q;h)$ for $h=2,3,4,6,$ and 8 (columns) and for $L=8$ (upper row) and $L=16$ (lower row). The gray curves are the numerical zero-field distributions $P^L(q)$, the points with error bars are the numerical $P^L(q;h)$, and the colored curves are the theoretical predictions obtained from the ultrametric tree (T.).}
    \label{fig:Lcomparison_straight_multifield}
\end{figure*}

Figure~\ref{fig:Lcomparison_straight_multifield} shows results for $P^L(q;h)$. For $L=8$ and $h=2$, the theory closely reproduces the numerical distribution. Differences become visible at $h=4$, mainly on the negative-$q$ side and around the positive-overlap peak. They are clearer for $h=6$ and 8, where the theory no longer reproduces the full shape of the numerical curve. Increasing the size from $L=8$ to 16 significantly improves the comparison, obtaining statistically compatible results up to $h=6$ and only finding small deviations for $h=8$.

\begin{figure*}
    \centering
    \includegraphics[width=0.87\textwidth]{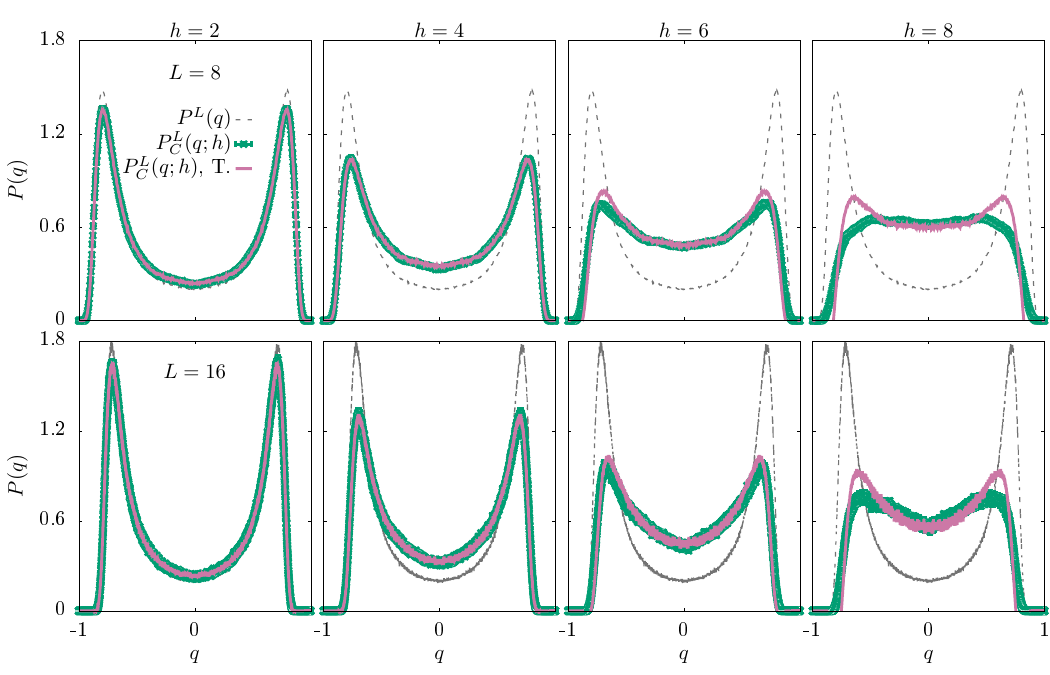}
    \caption{Comparison of $P_\text{C}^L(q;h)$ for $h=2,3,4,6,$ and 8 (columns) and for $L=8$ (upper row) and $L=16$ (lower row). The gray curves are the numerical zero-field distributions $P^L(q)$, the points with error bars are the numerical $P_\text{C}^L(q;h)$, and the colored curves are the theoretical predictions obtained from the ultrametric tree (T.).}
    \label{fig:Lcomparison_chaos_multifield}
\end{figure*}

The chaos comparison in Fig.~\ref{fig:Lcomparison_chaos_multifield} follows the same overall pattern. The agreement is excellent at $h=2$ and at $h=4$. At $h=6$ the differences become appreciable, especially for $L=8$, and at $h=8$ the theoretical curve systematically misses part of the numerical profile for both sizes. The comparison of the two rows again indicates that finite-size effects account for part, but not all, of the disagreement at strong fields. Taken together, the straight and chaos results support the interpretation that the reweighting prediction works in the weak-field regime and progressively loses accuracy once the perturbation is large enough to modify the states themselves.

\begin{figure*}[]
    \centering
    \includegraphics[width=\textwidth]{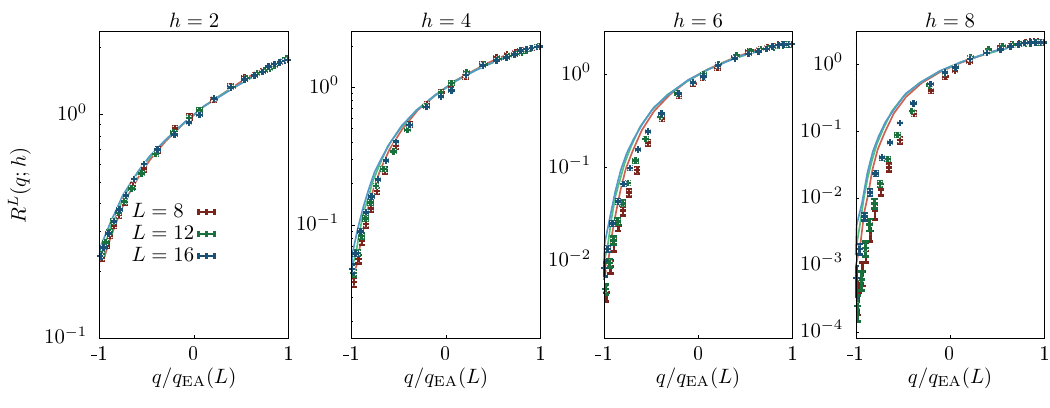}
    \caption{The straight ratio $R^L(q;h)$ versus $q/q_{\mathrm{EA}}(L)$ for $h=2,4,6,$ and 8 (columns) and $L=8,12,$ and 16 (colors). Points with vertical error bars are the numerical results, and solid lines are the analytical predictions from the ultrametric tree. The $L=8$ uncertainties are obtained by jackknife, whereas those for $L=12$ and 16 are propagated from the aggregate $P(q)$ and calibrated against the $L=8$ jackknife errors.}
    \label{fig:rstraight_multifield}
\end{figure*}

\begin{figure*}[]
    \centering
    \includegraphics[width=\textwidth]{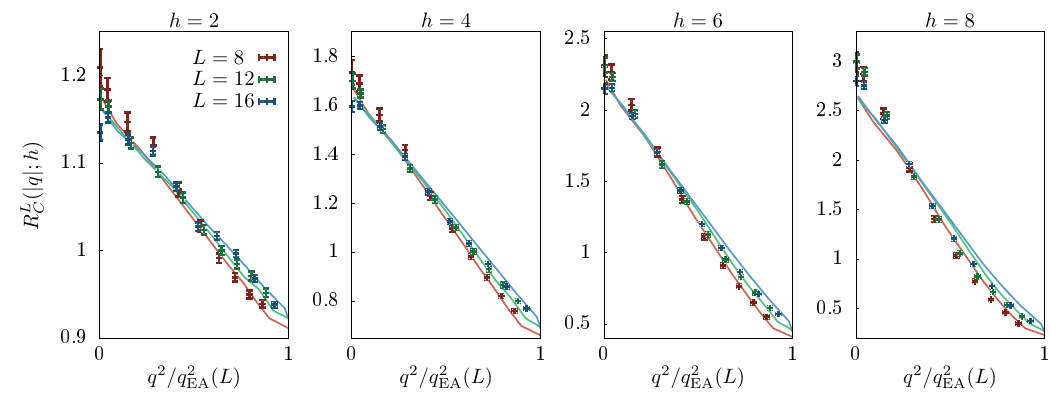}
    \caption{The chaos ratio $R_\text{C}^L(|q|;h)$ versus $q^2/q_{\mathrm{EA}}^2(L)$ for $h=2,4,6,$ and 8 (columns) and $L=8,12,$ and 16 (colors). Points with vertical error bars are the numerical results, and solid lines are the theoretical predictions from the ultrametric tree. The $L=8$ uncertainties are obtained by jackknife, whereas those for $L=12$ and 16 are propagated from the aggregate $P(q)$ and calibrated against the $L=8$ jackknife errors.}
    \label{fig:rchaos_multifield}
\end{figure*}

A similar conclusion can be reached by analyzing the $R^L$'s functions. In Figs.~\ref{fig:rstraight_multifield}, and~\ref{fig:rchaos_multifield} one can see the behavior of $R^L(q,h)$ as a function of $q/q_\mathrm{EA}$, and $R_\text{C}^L(q,h)$ as a function of $q^2/q^2_\mathrm{EA}$ for the three lattice sizes simulated~\footnote{Because the sample-resolved data were lost during the final stages of this work, the error bars for $L=12$ and $16$ in Figs.~\ref{fig:rstraight_multifield} and \ref{fig:rchaos_multifield} could not be estimated using the jackknife procedure employed for the $L=8$ data and for all other error bars in the paper. Instead, we obtained them by Gaussian resampling, propagating the uncertainty in the mean $P(q)$. Because this procedure systematically underestimates the uncertainties, we rescaled the resulting errors using correction factors determined, for each value of $h$, by comparing the propagated and jackknife errors at $L=8$. For $h=2,4,6,$ and $8$, respectively, these factors are $2.51$, $3.36$, $3.69$, and $3.82$ for the straight protocol, and $3.10$, $5.65$, $7.11$, and $8.03$ for the chaos protocol. Consequently, the error bars for $L=12$ and $16$ in these figures should be regarded only as order-of-magnitude estimates.}.
\newpage

\end{document}